\documentclass[conference]{IEEEtran}
\usepackage[dvipsnames]{xcolor}

\usepackage{hyperref}

\usepackage{booktabs} % For formal tables
\usepackage{subfig}
\usepackage{filecontents}
\usepackage{graphicx}
\usepackage{verbatim}
\usepackage{tikz}
\usepackage{algorithm}
\usepackage[noend]{algpseudocode}
\usepackage{multirow}
\usepackage{multicol, blindtext}
\usepackage{float}
\usepackage{amssymb}
\usepackage{gensymb}
\usepackage{textcomp}
\usepackage{verbatim}
\usepackage{bm}
\usepackage{url}

\usepackage{breakurl}
\usepackage{cite}
\usepackage{siunitx}

\usepackage{nomencl}

\ifCLASSINFOpdf
\else
\fi

\usepackage[cmex10]{amsmath}

\begin{document}
\graphicspath{{Figure/}}
\title{Hold the Door! Fingerprinting Your Car Key\\ to Prevent Keyless Entry Car Theft}


\author{\IEEEauthorblockN{Kyungho Joo*}
\IEEEauthorblockA{Korea University\\
khjoo0512@gmail.com}
\and
\IEEEauthorblockN{Wonsuk Choi*}
\IEEEauthorblockA{Korea University\\
wonsuk85.choi@gmail.com}
\and
\IEEEauthorblockN{Dong Hoon Lee}
\IEEEauthorblockA{Korea University\\
donghlee@korea.ac.kr}}

\IEEEoverridecommandlockouts
\makeatletter\def\@IEEEpubidpullup{6.5\baselineskip}\makeatother
\IEEEpubid{\parbox{\columnwidth}{
    * Co-first Authors\\
    Network and Distributed Systems Security (NDSS) Symposium 2020\\
    23-26 February 2020, San Diego, CA, USA\\
    ISBN 1-891562-61-4\\
    https://dx.doi.org/10.14722/ndss.2020.23107\\
    www.ndss-symposium.org
}
\hspace{\columnsep}\makebox[\columnwidth]{}}

\maketitle

\begin{abstract}
%-------------------------------------------------------------------------------
Recently, the traditional way to unlock car doors has been replaced with a keyless entry system which proves more convenient for automobile owners. When a driver with a key fob is in the vicinity of the vehicle, doors automatically unlock on user command. However, unfortunately, it has been shown that these keyless entry systems are vulnerable to signal-relaying attacks. While it is evident that automobile manufacturers incorporate preventative methods to secure these keyless entry systems, they continue to be vulnerable to a range of attacks. \textcolor{black}{Relayed signals result in valid packets that are verified as legitimate, and this makes it is difficult to distinguish a legitimate door unlock request from a malicious signal.}
In response to this vulnerability, this paper presents an RF-fingerprinting method (coined ``HOld the DOoR", \texttt{HODOR}) to detect attacks on keyless entry systems - the first attempt to exploit the RF-fingerprint technique in the automotive domain.
\texttt{HODOR} is designed as a sub-authentication method that supports existing authentication systems for keyless entry systems and does not require any modification of the main system to perform. 
Through a series of experiments, the results demonstrate that \texttt{HODOR} competently and reliably detects attacks on keyless entry systems. 
\texttt{HODOR} achieves both an average false positive rate (FPR) of $0.27\%$ with a false negative rate (FNR) of $0\%$ for the detection of simulated attacks, corresponding to current research on keyless entry car theft.
Furthermore, \texttt{HODOR} was also observed under environmental factors: temperature variation, non-line-of-sight (NLoS) conditions, and battery aging.
\texttt{HODOR} yields a false positive rate of $1.32\%$ for the identification of a legitimated key fob even under NLoS conditions.
Based on the experimental results, it is expected that \texttt{HODOR} will provide a secure service for keyless entry systems, while remaining convenient.
\end{abstract}

\section{Introduction}
%-------------------------------------------------------------------------------
Recently, keyless entry systems have been developed and installed in modern vehicles for the convenience of drivers. Before the keyless entry system, it was necessary to physically insert a key into the key hole to unlock the doors of a vehicle. This traditional way to unlock doors was inconvenient as well as vulnerable to physical key copying leading to relatively easy automotive theft or break-ins. The keyless entry system enables a driver to unlock doors without inserting anything, via two distinct systems: the remote keyless entry (RKE) system and the passive keyless entry and start (PKES) system. The RKE system unlocks doors with the press of a button on a remote key fob at a distance. In the PKES system, car doors are automatically unlocked as the user makes physical contact with a button on a door when the key fob is in the vicinity. This implies that drivers no longer need to remove their key fobs from their pockets or bags. We note that the PKES system is mostly designed to include the remote lock and unlock functions provided by the RKE system. However, as keyless entry systems are becoming commonplace on modern vehicles, cyber security attacks are also on the rise. Vehicle manufacturers, therefore, have applied their own security mechanisms to verify either the remotes or key fobs. In particular, encryption with a pre-shared, long-term secret key and rolling codes \cite{waraksa1995rolling, farris2000rolling} are common methods used to verify a legitimate key fob. \par
\textcolor{black}{Despite these security mechanisms, several vulnerabilities with keyless entry systems have been discovered. In 2010, the authors of \cite{francillon2011relay} demonstrated a relay attack on PKES systems, in which vehicle doors were unlocked. In the relay attack, two colluding adversaries would work in concert to extend the original range of RF communication between a vehicle and its key fob. One adversary must be close to the target vehicle and the other must be close to its key fob. They cooperate with each other to relay signals from the vehicle to the key fob side. As a consequence, even outside of the pre-defined communication range, the vehicle and its key fob interact with each other, which leads to the unlocking of the doors.} In Germany and the United Kingdom, automotive thieves successfully carried out these types of signal-relaying attacks, which were captured on security cameras \cite{Keylesss80:online, Stolenca94:online}. In addition, an adversary could exploit a particular vulnerability of a cryptographic algorithm used in the remote keyless entry system to extract a pre-shared secret key between the vehicle and its key fob, thereby creating and transmitting a malicious message for a door unlock command \cite{garcia2016lock, verdult2012gone, immler2012breaking, benadjila2017one, hicks2018dismantling}. Furthermore, recent studies have shown that long-term secrets can be compromised not only in the RKE system but also in the PKES system \cite{wouters2019fast}.\par
The underlying reason of the cyber security attacks on the keyless entry system is that radio frequency (RF) signals emitted from key fobs can be relayed or replayed regardless of active security methods like encryption or authentication. Since the keyless entry system accepts any request for authentication as long as valid signals are within the communication range, extension of the communication range by relaying or forwarding a signal ultimately enables an attacker to unlock car doors. \textcolor{black}{One approach to resolve this issue might be the use of an RF distance-bounding protocol that verifies the actual physical proximity of a request \cite{brands1993distance}, \cite{hancke2005rfid}. However, RF distance-bounding protocols are highly sensitive to timing errors. This is because the distance-bounding protocol measures the distance based on the time of flight (ToF) of an RF signal which propagates at the speed of light. Recently, an ultra-wide band impulse radio (UWB-IR) ranging technique has emerged as a prominent technology to deploy a distance-bounding protocol, and numerous efforts are underway to deploy a secure UWB-IR ranging technique \cite{IEEE802157:online, MTAC2020, singh2019uwb, singh2017uwb}. However, this approach would require the keyless entry system to adopt an entirely new communication system to implement the RF distance-bounding protocol.}
\par
To detect attacks on keyless entry systems, we employ an RF fingerprinting technique that extracts fingerprints of individual RF devices from their RF signals. Due to hardware imperfections, distinct characteristics per RF device can be extracted even if they transmit the same binary message. In other areas, RF fingerprinting methods have already been proposed to identify RF devices \cite{brik2008wireless, danev2009transient, padilla2013radiofrequency, danev2009physical, zanetti2010physical}, which are referred to as the ground truth of \texttt{HODOR}. These existing methods were designed to identify RF devices in line-of-sight (LoS) and indoor conditions. However, \texttt{HODOR} is herein proven to function in both non-line-of-sight (NLoS) and outdoor conditions.\par
In this paper, we present our evaluations of \texttt{HODOR} in detecting attacks on keyless entry systems, including RKE and PKES systems. The method has been designed as a sub-authentication system that supports an existing authentication system. As such, it can be directly applied to a keyless entry system without any modification to the current communication system. Our experimental results show that \texttt{HODOR} precisely and accurately detects several types of attack attempts. The detailed contributions of this research are as follows.
\begin{itemize}
\item \textcolor{black}{Based on previous attack demonstrations conducted in existing research, we present a new attack model that combines all known attack methods; our attack model covers both PKES and RKE systems. This is the first attempt to formalize existing attacks on keyless entry systems.}
\item We present an RF fingerprinting method, \texttt{HODOR}, to identify legitimate key fobs and detect malicious attempts defined in our attack model. \texttt{HODOR} can be easily employed by adding a new device that captures and analyzes the ultra-high frequency (UHF) band RF signals emitted from a key fob, which implies that the current system would require no alterations.
\item We performed a series of experiments to evaluate \texttt{HODOR}. We simulated malicious attacks that are defined in our attack model with different kinds of RF devices. The experimental results show that \texttt{HODOR} is able to correctly detect the attacks, which has an average FPR of 0.27$\%$ with an FNR of 0$\%$ for the PKES system.
\item \texttt{HODOR} was also evaluated under varying environmental factors, such as temperature variations, NLoS conditions (e.g., a key fob placed in a pocket) and battery aging. We show that the features we present work properly under these environmental factors, indicating that \texttt{HODOR} can be applied in current commonly existing systems.
\end{itemize}
%\vspace*{-0.5cm}
%-------------------------------------------------------------------------------
\section{Background}
%-------------------------------------------------------------------------------
%\vspace*{-0.2cm}
To easily understand \texttt{HODOR} and our attack model, we describe the background of keyless entry systems and digital communication.
\subsection{Keyless Entry Systems}
\label{Subsection:Keyless_entry_system}

\begin{figure}[t]
\centering
\subfloat[Message flow of PKES system]{%
  \includegraphics[clip,width=0.9\columnwidth]{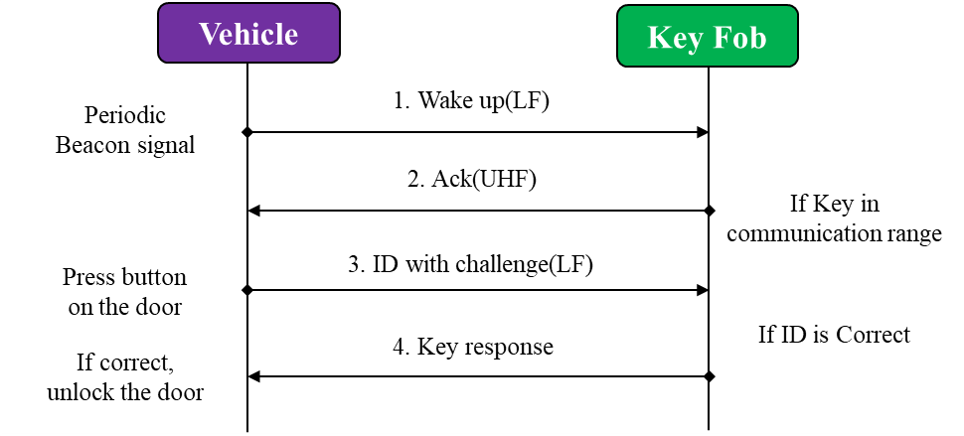}%
  \label{fig:MF_PKE}
}

\subfloat[Message flow of RKE system]{%
  \includegraphics[clip,width=0.9\columnwidth]{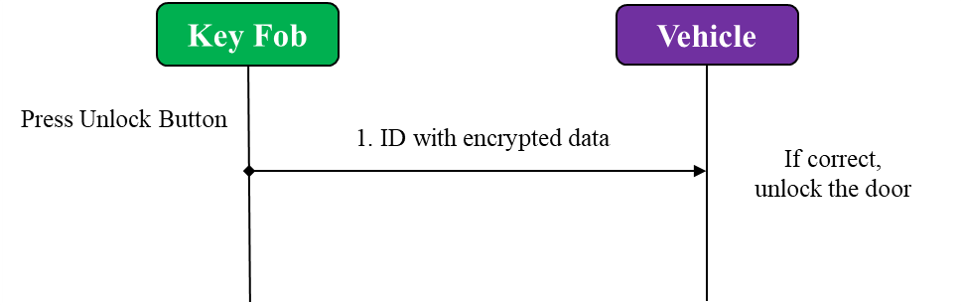}%
  \label{fig:MF_RKE}
}
\label{fig:MF}
\caption{Message flow of each keyless entry system}
\end{figure}

When a vehicle is equipped with a PKES system, a driver can unlock their doors as long as he/she is in the vicinity of their vehicle. In some cases, a driver must also press a button on the vehicle. Mutual communication between the key fob and vehicle is needed to verify whether the driver is actually nearby. The vehicle then sends a challenge to the key fob via low frequency (LF) band (125 $\sim$ 135kHz) communication, and the corresponding key fob responds to the request via UHF-band communication. Fig. \ref{fig:MF_PKE} illustrates an example of the message flow for verification in the PKES system. Vehicles periodically broadcast LF signals to check if a proper key fob is in the vicinity, like beacon signals \cite{gerasenko2001beacon}. In the case that a key fob is within the communication range of an LF band (e.g., 1 $\sim$ 2m), it receives a periodical LF-band signal from the vehicle which enables it to transmit the response signal in a UHF band. In North America, 315MHz is assigned for the UHF-band, whereas 433.92MHz or 868MHz is assigned in Europe \cite{garcia2016lock}. For security reasons, packets are encrypted with a long-term secret key that is shared between a remote key fob and its corresponding vehicle in advance. It is noted that in the PKES system, a driver is even able to start the engine without inserting a physical key into the ignition switch.\par
In an RKE system, UHF-band signals from the key fob are transmitted unidirectionally. Only when the driver \textcolor{black}{presses} the button on the key fob is the UHF-band signal transmitted. Fig. \ref{fig:MF_RKE} illustrates an example of the messages flow for verification in the RKE system. The transmission of the same RF signals from a remote key fob is repeated multiple times to increase the reliability of communication \cite{garcia2016lock}.

%\vspace*{-0.2cm}
\subsection{Digital Communication}
\textcolor{black}{The information source (i.e., binary code) is encoded to be delivered via wireless communication. Next, encoded binary information is mapped into a \textit{symbol} and conveyed to an analog RF signal (the so-called baseband signal) through various modulation schemes. Frequency shift keying (FSK) and amplitude shift keying (ASK) modulation are the most common methods for modulating RF signals in keyless entry systems. Specifically, \textit{symbol} consists of several bits. But in the case of Binary-ASK and Binary-FSK where a \textit{symbol} consists of a single bit, the meaning can be seen as interchangeable. Since most keyless entry systems use BFSK or BASK, for the remainder of this paper, we will use the terms interchangeably.
}\par

\begin{figure}[t]
    \centering
    \includegraphics[clip,width=0.85\columnwidth]{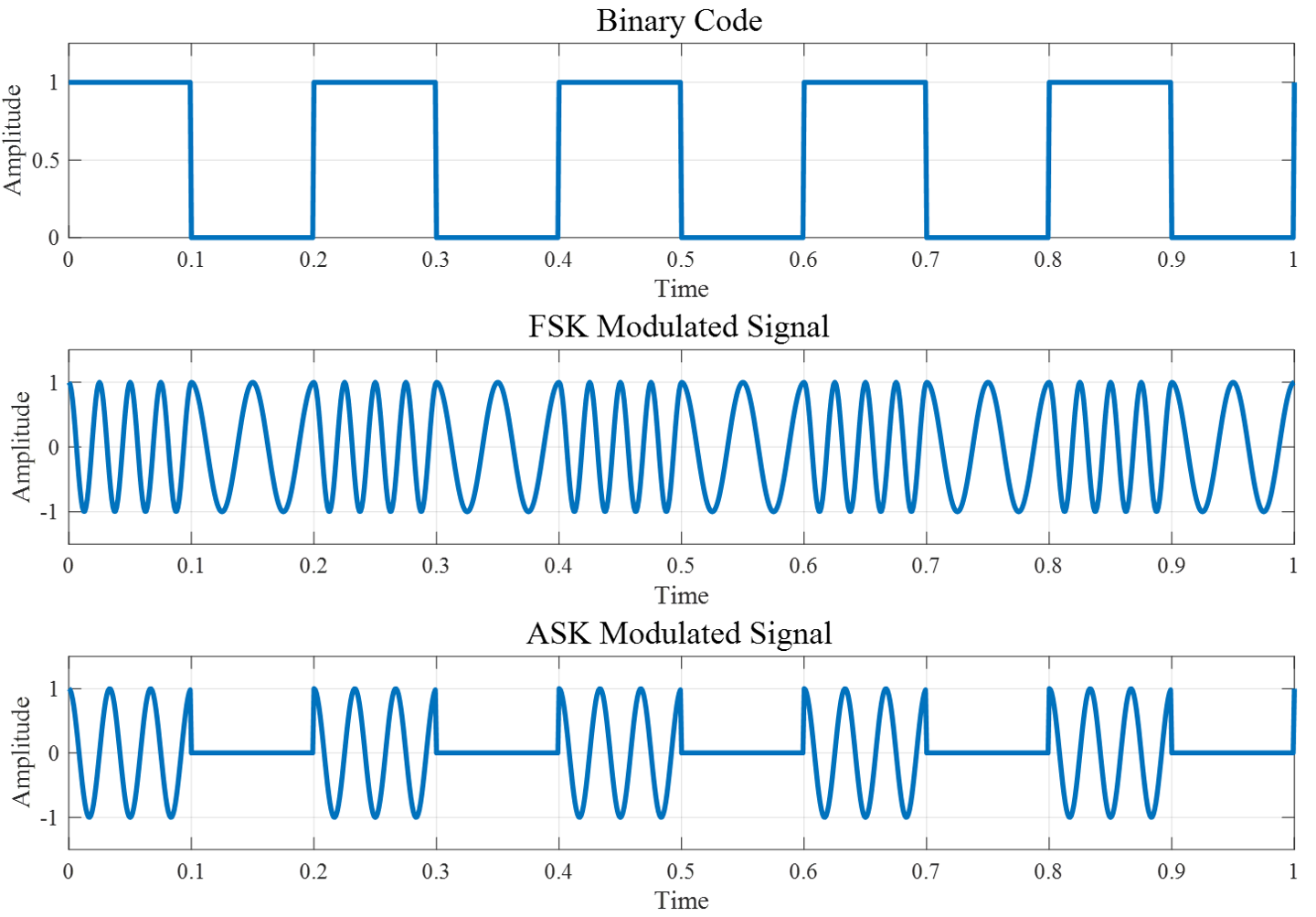}%
    \caption{An example of FSK and ASK modulation}
    \label{fig:ASK_FSK_Eg}
\end{figure}

\textcolor{black}{
BFSK expresses bits 0 and 1 by shifting the frequency of the baseband signal by a specific frequency deviation based on the center frequency $f_c$. The center frequency $f_c$ means the frequency of the carrier signal. A carrier signal is a sinusoidal wave with a carrier frequency. This carrier frequency is defined as standard according to its application and region. That is, $f_c$ of a UHF-band signal is 315MHz or 433.92MHz and $f_c$ of an LF-band signal is 125 $\sim$ 135Khz. Fig. \ref{fig:ASK_FSK_Eg} shows an example of one FSK and one ASK modulated signal corresponding to the binary code. In FSK modulation, if a sinusoidal wave of the frequency $f_{c}+f_{d}$, which is higher than the center frequency by $f_{d}$ indicates bit 1, a sinusoidal wave having a frequency of $f_{c}-f_{d}$, which is lower than the center frequency by $f_{d}$, indicates bit 0. The $f_{d}$ indicates the frequency deviation. BASK expresses 0 and 1 bits using the amplitude of sinusoidal with center frequency $f_c$. If a sine wave having an amplitude of $A_0$ indicates bit 1, a sine wave having an amplitude of 0 indicates bit 0. Finally, the baseband signal is mixed with a carrier signal and transmitted through an antenna. Since a high-frequency electric signal is emitted from the antenna, the RF signal can be physically transmitted through the air in the form of an electromagnetic wave. At the receiver side, the received signal is processed through a mixer, demodulator, and decoder - reverse order of the transmission process. We refer readers to \cite{sklar1988digital} for further reading on digital communication.}

%-------------------------------------------------------------------------------
\section{System Model}
%-------------------------------------------------------------------------------
This section presents the system overview, including how \texttt{HODOR} detects malicious attacks on keyless entry systems. For a clear understanding of \texttt{HODOR}, we define the attack model that simulates the actions of adversaries and their capabilities.

\subsection{System Overview}
\begin{figure}[t]
\centering
  \includegraphics[clip,width=1\columnwidth]{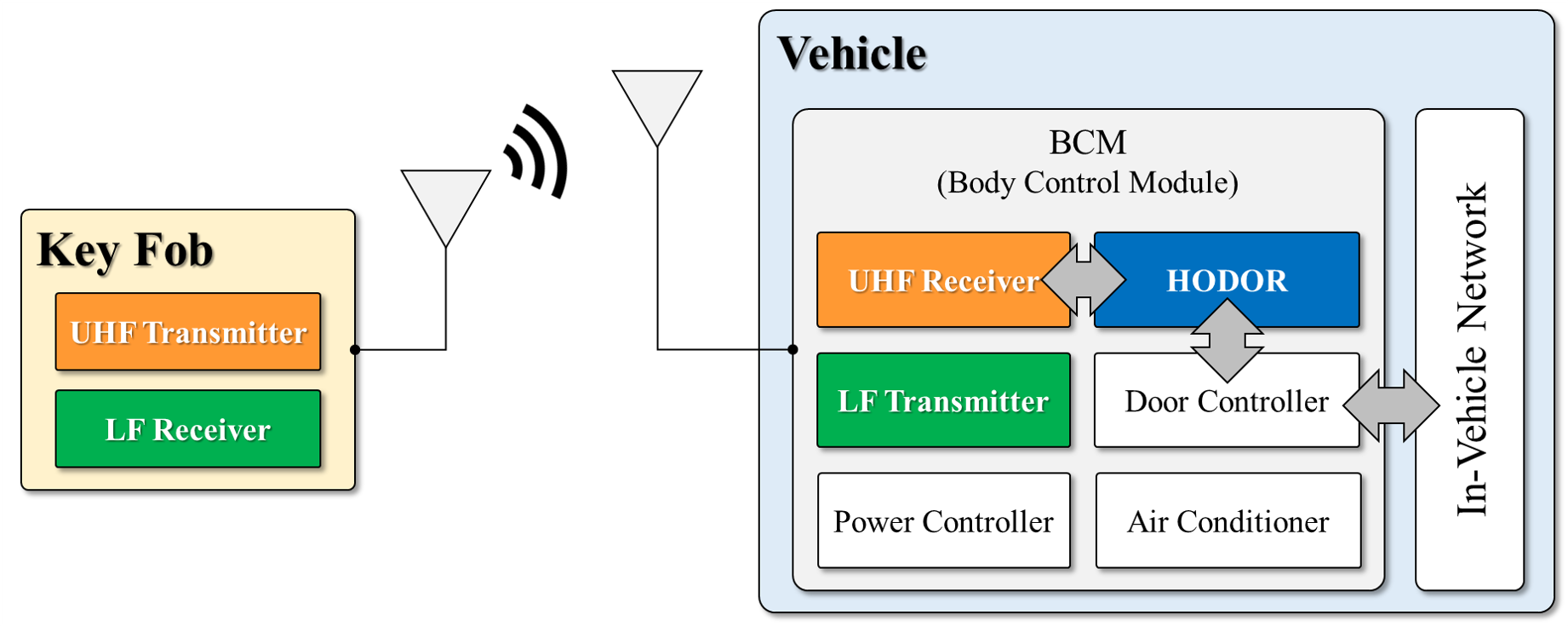}%
  \caption{System model}
  \label{fig:SysModel}
\end{figure}

\textcolor{black}{In this subsection, we present the system overview of \texttt{HODOR}. The vehicle should verify the UHF-band signals emitted from the key fob. Therefore, \texttt{HODOR} should be equipped with an RF receiver and mounted to the vehicle and integrated with the Body Control Module (BCM) of a car which controls various electronic accessories in the car's body. One typical function of BCM is transmitting a lock/unlock command packet through the in-vehicle network communication such as CAN or LIN. In the case of an attack being detected, \texttt{HODOR} raises an attack detection alarm and BCM does not transmit the CAN packet which contains the unlock command. Fig. \ref{fig:SysModel} illustrates \textcolor{black}{the} overall system model of \texttt{HODOR}.}
%\vspace*{-0.2cm}

%\vspace*{-0.2cm}
\subsection{Attack Models}
We present a new attack model for PKES systems. \textcolor{black}{Our attack model covers attacks on PKES systems, which were implemented with the LF/UHF band RFID communication. In addition, our attack model also covers existing demonstrations of attacks on RKE systems using UHF band RFID communication.} In our attack model, the main objective of a hypothetical adversary is to unlock a vehicle. For the simplicity of the attack model, we excluded other functions, such as an engine start message. In addition, physical damage to a vehicle is beyond the scope of our attack model, regardless of whether doors are successfully unlocked via a malicious message.
\textcolor{black}{Three different types of attacks on PKES systems were outlined in our attack model in regards to how an adversary might deliver a valid message that enables the agent to unlock doors. For a relay attack, we categorized these into \textbf{Single-band relay attack} and \textbf{Dual-band relay attack}. Two adversaries must collaborate to accomplish either a single-band or dual-band relay attack because relayed signals are used in this model to extend communication range. In a \textbf{Cryptographic attack}, however, there is a single adversary who attempts to unlock doors on the keyless entry system. In addition to the attack model on the PKES system, we also considered an attack model for the RKE system. In \textbf{Attacks on RKE system}, we demonstrate that all known attacks on RKE systems can be grouped into the two categories.} \par
\label{subsec:attack model}
%\vspace*{-0.2cm}
\subsubsection{\textbf{Single-band relay attacks}} In the PKES system, a vehicle transmits a verification request to the corresponding key fob using the LF-band RFID communication. When the request is received in the LF band, the key fob automatically responds in the UHF band. The PKES system was originally intended to only operate within the LF-band communication range (e.g., 1 $\sim$ 2 meters). However, by relaying an LF-band request from a vehicle to its key fob, an adversary is able to make a key fob respond to a request within the UHF-band communication range (e.g., up to 200 meters) even if it is out of the LF-band communication range. It should be noted that the communication range of a key fob differs per manufacturer. In essence, a single-band relay attack aims to relay an LF-band request to the key fob, in which case the UHF-band response for the LF-band request is directly transmitted to the vehicle. In other words, the UHF-band response is not relayed and is transmitted within its communication range. This paper focuses on the UHF-band RF signals emitted from key fobs, and, as such, does not consider how the LF-band signals are relayed to the key fobs.\par\noindent
%\vspace*{-0.5cm}
\subsubsection{\textbf{Dual-band relay attacks}} Unlike a single-band relay attack, a dual-band relay attack is not only able to relay a request from the vehicle within the LF band, but also a response from a key fob within the UHF band. Accordingly, in a dual-band relay attack, the PKES system can be attacked even if the key fob is much farther away from the vehicle during a single-band relay attack. Adversaries intending to commit a dual-band attack must also possess industry-standard equipment in order to relay both the LF-band and the UHF-band RF signals. The UHF-band signals can be delivered to the vehicle by a signal-extending module \cite{Keylesss80:online, Stolenca94:online}, or through two adversaries, one who would record and forward the UHF-band signal out of communication range and the other who would inject the forwarded signal into the vehicle \cite{unicornteam:online}. We denote the former as an  \textbf{Amplification attack} and the latter as a \textbf{Digital relay attack}. The difference between the two attack types is whether the adversaries perform digital communication to forward binary information contained in LF/UHF-band signals.\par
During an \textbf{Amplification attack}, adversaries simply amplify both the LF band and the UHF-band signals using the RF amplifier. There are two ways to inject UHF-band signals to the vehicle. First, the adversary at the key fob side amplifies the UHF-band signals and directly injects it into the vehicle. Second, both adversaries amplify the UHF-band signals. Although the latter case can produce a higher signal strength than the former, the RF amplifier intensifies both the pass-band signal and the noise leading to unintended feature variation. Therefore, in Section \ref{sec:eval}, we have simulated an amplification attack based on the former case.
In a \textbf{Digital relay attack}, \textcolor{black}{adversaries perform the} whole process of digital communication to inject an attack signal. Adversaries demodulate and decode the LF/UHF-band signal to forward binary information to each other. The delivery of binary information can be conducted through various wireless communication systems such as Wi-Fi or Bluetooth. After receiving the binary information, the adversary injects an attack signal through an encoding and modulation process. The advantage of a digital relay attack is that the communication range can be much larger than with a single-band relay attack or amplification attack. This is because binary information is forwarded through state-of-the-art digital communication. However, since most PKES systems assign a maximum delay \cite{francillon2011relay}, the attack signal should be injected within the maximum delay period. Nevertheless, researchers have shown that digital relay attacks can be successfully mounted with cheap RF devices \cite{unicornteam:online}.
%\vspace*{-0.4cm}
\subsubsection{\textbf{Cryptographic attacks}} An adversary can extend the communication range between a vehicle and its key fob, as well as mount a cryptographic attack. In a cryptographic attack, the adversary exploits the weaknesses of the cryptographic algorithm which is equipped in the PKES system. In the vicinity of the key fob, the adversary injects malicious LF-band signals (challenge) to the key fob and collects the UHF-band signals (response). Due to the lack of mutual authentication in the PKES system, the key fob accepts malicious LF-band signals and transmits corresponding responses. After collecting sufficient challenge and response pairs, the adversary performs cryptanalysis to extract a long-term secret key. Consequently, the adversary can inject valid UHF-band signals depending on a challenge signal from the vehicle. A 2018 study has shown that the PKES system of Telsa Model S is equipped with a weak cryptographic algorithm and does not require mutual authentication \cite{wouters2019fast}. Researchers have uncovered that the outdated proprietary cipher DST40 has been mounted to the Telsa Model S.
\textcolor{black}{Furthermore, unrevealed PKES systems with weak cryptographic algorithms or key management failure \cite{garcia2016lock} are also expected to be vulnerable to a cryptograhic attack, given that the adversary extracts the binary code and injects the attack signal. With regards to \texttt{HODOR}, this attack scenario is considered to be the same as the transmitted signal that would be analyzed in a digital relay attack.}
\subsubsection{\textbf{Attacks on RKE systems}}\textcolor{black}{We categorized the attacks on PKES systems as Single-Band Relay, Dual-Band Relay, and Cryptographic Attack. In addition, previous studies have shown that where a long-term secret key is used in an RKE system, it can be compromised by an adversary through the use of cryptanalysis with reverse engineering \cite{benadjila2017one, BogdanovLinearSlideAttacks, verdult2012gone, hicks2018dismantling}, an exhaustive key search \cite{verstegen2018hitag, immler2012breaking}, or combining both methods \cite{garcia2016lock}. As a result, an adversary can generate a valid packet in a similar manner to a cryptographic attack. To the best of our knowledge, our cryptographic attack model also covers all known attacks on RKE systems except a \textit{rolljam attack}. In a \textit{rolljam attack}, an adversary performs a jamming attack and eavesdrops on valid UHF signals. When the driver presses the unlock button on the key fob, the vehicle remains locked because the signal has been blocked by the jamming attack, and the driver will naturally attempt to unlock the door again. This creates a second signal that is also recorded and blocked, however, at this time, the adversary replays the first code to unlock the door. As a result, the driver assumes that the key fob is working normally. However, the adversary can now inject an attack signal using a second rolling code which has not been received by the vehicle.}

%\vspace*{-0.25cm}
%-------------------------------------------------------------------------------
\section{Our Method: \texttt{HODOR}}
%-------------------------------------------------------------------------------
\label{sec:method}
\subsection{Overview}
\begin{figure}[!t]
\centering %% Check the Normalization Parameter Generation
  \includegraphics[clip,width=0.85\columnwidth]{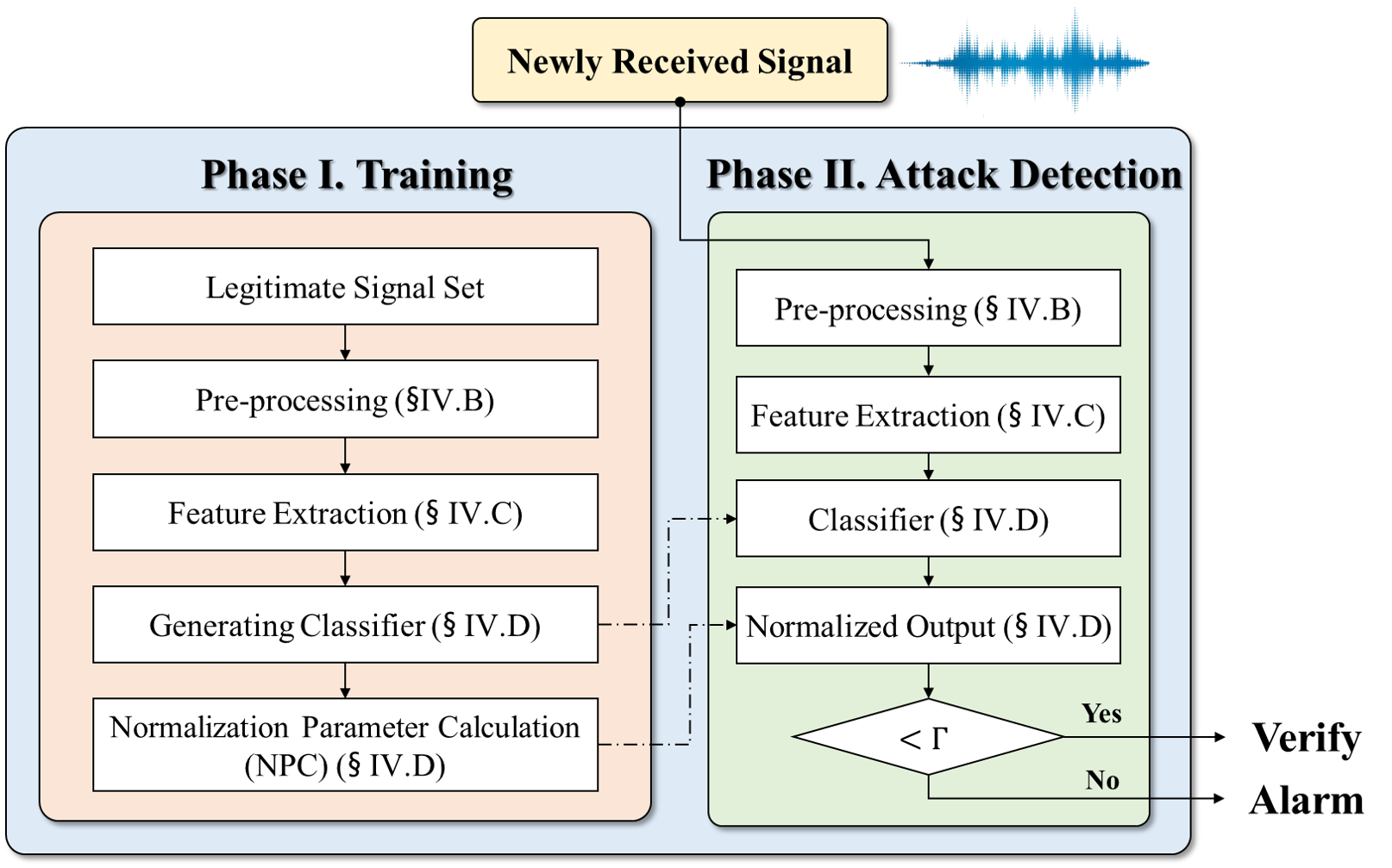}%
    \caption{Overview of \texttt{HODOR} architecture}
    \label{fig:hodor_overview}
\end{figure}

\textcolor{black}{In this section, we explain our design decisions to realize \texttt{HODOR}. Fig. \ref{fig:hodor_overview} shows an overview of \texttt{HODOR}'s architecture. \texttt{HODOR} aims at detecting an attack signal using a classifier which is generated by legitimate signals only. There are two main phases in \texttt{HODOR}: the \textit{Training} phase and the \textit{Attack Detection} phase. In the \textit{Training} phase, \texttt{HODOR} creates a classifier based on a training dataset which contains only legitimate signals. Through preprocessing and feature extraction, a set of features per RF signal are obtained and the classifier is trained. In addition, normalization parameters, which are used for output normalization in the \textit{Attack Detection} phase, are computed. After the classifier is trained, in the \textit{Attack Detection} phase, \texttt{HODOR} is now able to detect any attacks defined in our attack model in Section \ref{subsec:attack model}. In the \textit{Attack Detection} phase, \texttt{HODOR} receives a new RF signal which contains a door unlock request. Then, \texttt{HODOR} conducts preprocessing and feature extraction on this newly received RF signal, as outlined in the \textit{Training} phase. The extracted feature set is used as input to the trained classifier, and \texttt{HODOR} makes a decision whether the received RF signal has been transmitted from a legitimate key fob or not. This decision is made based on the normalized output of the classifier and a pre-defined threshold. In an invalid case, when the normalized output is larger than the threshold, the corresponding door unlock request is not validated and \texttt{HODOR} alerts the BCM module.}

\subsection{Preprocessing}
\begin{figure}[!t]
\centering
  \includegraphics[clip,width=\columnwidth]{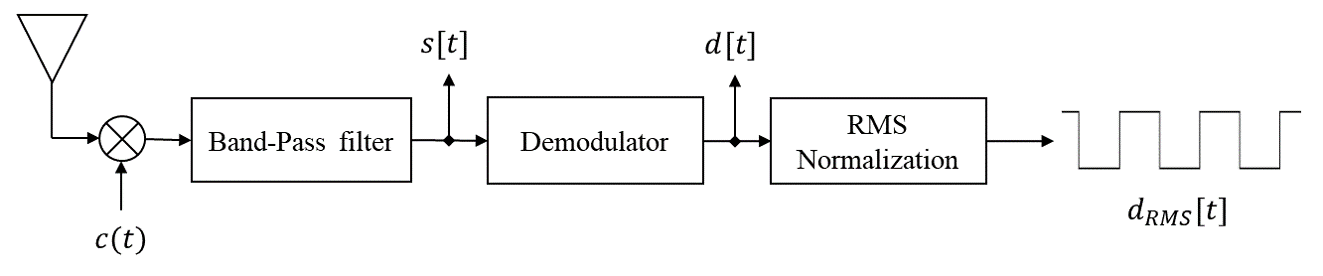}%
    \caption{Preprocessing block diagram}
    \label{fig:Blockdiagram}
\end{figure}

At the outset, \texttt{HODOR} receives UHF-band RF signals, which become preprocessed as illustrated in Fig. \ref{fig:Blockdiagram}. The received signal, including a carrier signal $c(t)$, a baseband signal $s(t)$, and channel noise $n(t)$, is denoted as follows.\\
\begin{equation}
r(t) = s(t)\otimes c(t)+n(t)
\end{equation}
where $\otimes$ is the operation for the mixer. The carrier signal is a sinusoidal signal at the carrier frequency ($f_c$) of the UHF band. To obtain meaningful information for analysis located in the baseband signal, the carrier signal must be removed. In other words, the received raw signal is shifted back down to the baseband by mixing the sinusoidal signal at the same carrier frequency as follows. \\
 \begin{equation}
 \label{eq_demo}
 r[t]\otimes c[t] = s[t]+n[t]\otimes c[t]
 \end{equation}
It should be noted that \texttt{HODOR} samples a continuous analog RF signal, and owing to this, we denote the sampled signal as $[t]$ which represents discrete values.
To remove $n[t]\otimes c[t]$, the bandpass filter is performed on $r[t]\otimes c[t]$. As a result, we obtain the baseband signal $s[t]$ from the received signal $r[t]$.\par
Subsequently, \texttt{HODOR} demodulates the baseband signal $s[t]$ into a pulse signal $d[t]$. The pulse signal is encoded from a binary code. As mentioned in Section \ref{subsec:attack model}, FSK and ASK are typical modulation schemes used in keyless entry systems, which are determined by manufacturers. After demodulation, the pulse signal is normalized to scale its power to a certain value. Since the received signal strength (RSS) is determined by a channel condition, it would be difficult to reliably extract the features under noisy channel conditions. To be independent to the degree of RSS, \texttt{HODOR} applies root-mean-square (RMS) normalization, through which the power of a demodulated signal is scaled as 1. For example, if $d[t]$ is composed of $N$ samples ($d_1$, $d_2$, ..., $d_N$), the RMS-normalized signal is calculated as follows.\\
\begin{equation}d_{RMS}[t] = \frac{d[t]}{\sqrt{\frac{\sum_{i=1}^{N}\ d_{i}^{2}}{N}}}\end{equation}

\subsection{Feature Extraction}
On the preprocessed signals, \texttt{HODOR} extracts salient features by which a legitimate request and a malicious attempt are distinct. In wireless transmissions, the radio preamble (sometimes called a header) is used to synchronize the clock between a transceiver and a receiver. Preamble has a static bit sequence independent of the data packet. \texttt{HODOR} extracts the features from the preamble of the pulse signal since it allows \texttt{HODOR} to extract features independent of the data and the key fob. We propose four types of features: \romannumeral 1) peak frequency, \romannumeral 2) frequency offset, \romannumeral 3) SNR, and \romannumeral 4) a set of statistical features.

\par
\begin{figure}[t]
\centering
\subfloat[]{%
  \includegraphics[clip,width=0.44\columnwidth]{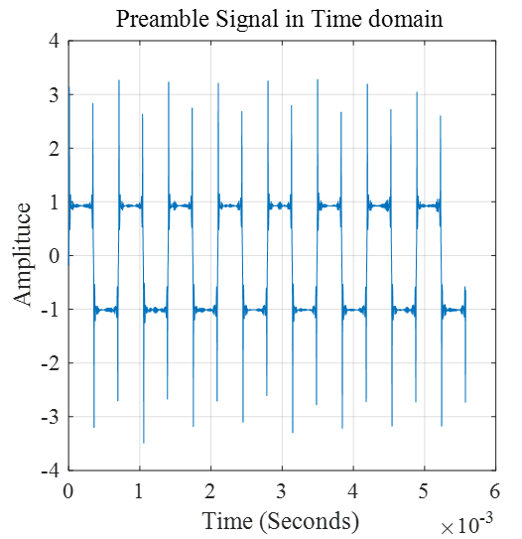}%
  \label{fig:Preamble signal in time domain}
}
\subfloat[]{%
  \includegraphics[clip,width=0.46\columnwidth]{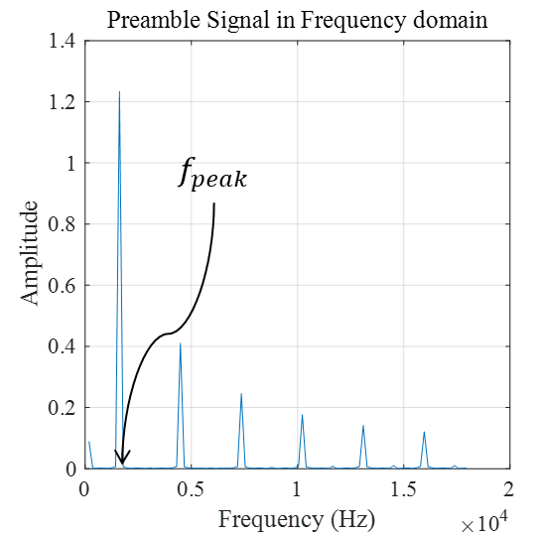}%
  \label{fig:Preamble signal in frequency domain}
}
\caption{Preamble signal in time and frequency domains: (a) Preamble signal in time domain, (b) Preamble signal in frequency domain}
\label{fig:Preamble signal}
\end{figure}

\noindent
\indent\textbf{Peak frequency. } Peak frequency is a feature in a frequency domain. Since the preamble part of a time-domain pulse signal is given, it must be transformed to the frequency-domain signal by fast fourier transform (FFT). 
Fig. \ref{fig:Preamble signal} shows the preamble part of the UHF-band RF signal transmitted from a key fob and its FFT result.
It can be seen that several dominant peaks exist in the frequency-domain preamble signal. 
The peak frequency ($f_{peak}$) is the frequency where the highest amplitude value exists as follows.
\begin{equation}
f_{peak} = \operatorname*{arg\,max}_f |D_{RMS}[f]|
\end{equation}
where $D_{RMS}[f]$ is the FFT result of $d_{RMS}[t]$.
The peak frequency feature is affected by a clock source used for micro controllers. Because of the imperfection of clock sources, different peak frequency values can be extracted from different RF devices. Accordingly, this feature is used to distinguish a legitimate key fob from other devices used for malicious attacks.\par\noindent

\indent\textbf{Carrier frequency offset. } \textcolor{black}{Carrier frequency offset is another feature in the frequency domain of a preamble signal. Different from \textit{peak frequency}, carrier frequency offset is the feature that is extracted from the baseband signal s[t] which is shown in Fig. \ref{fig:ASK_FSK_Eg}. As each key fob has a non-ideal (i.e., deviated from 433.92MHz) carrier frequency ($f_{c}'$) due to the hardware imperfection, the RF receiver works on a wide band to reliably receive signals from the key fob \cite{yang2018inside} in a real vehicle. This imperfection also occurs in the receiver, which leads the non-ideal frequency of $f_{c}''$ when generating $c(t)$. Consequently, when the $r(t)$ is mixed with $c(t)$, the baseband signal ($s[t]$) in Equation (2) has a different frequency offset value ($f_{c}'-f_{c}''$) according to each transmitter and receiver pair. We exploit this frequency difference as a feature to verify legitimate key fobs and denote it as $f_{c}^{offset}$.}

\indent\textbf{SNR. } Signal-to-noise ratio (SNR) is a measure that compares the level of a desired signal to the level of background noise. SNR is defined as the ratio of signal power to noise power. In addition, SNR is expressed in decibels (dB) as follows.
\begin{equation}
SNR_{dB} = 10\log_{10}\frac{P_{signal}}{P_{noise}}
\end{equation}
Where $P_{signal}$ is the power of a demodulated signal (i.e., meaningful information) and $P_{noise}$ is the power of background noise (i.e., unwanted signal). It is noted that measures greater than 0 dB indicate more signal than noise. 
In the PKES system, a UHF-band RF signal emitted at a larger distance than the proximity distance should be considered malicious even if the signal is from the legitimate key fob. Since a feature of SNR is easily affected by channel conditions, path loss, or the reduction in power density of a signal as it propagates through space, can be estimated. By analyzing the features of SNR, \texttt{HODOR} is able to verify whether a particular signal has been generated within a specific vicinity. \par\noindent
\indent\textbf{Statistical features. } \textcolor{black}{To support the above features, a set of statistical features were also employed. Statistical features represent the various characteristics of a sampled signal. As a result, numerous studies on signal processing and wireless communication area employed statistical features for node or channel identification \cite{das2014you, murvay2014source, dey2014accelprint}. Since the hardware characteristics of an attack device and channel condition affect the signal characteristics, we employed the statistical features to differentiate an attack signal. With the three crafted features and 20 statistical features used in \cite{das2014you}, we ran a feature selection algorithm to eliminate features that are not beneficial to performance. From there, we selected the top five highest performing features and decided to not employ all 23 while testing \texttt{HODOR} in consideration of execution time during the feature extraction phase. As more features are used, more time is required to compute the features, and this time delay hinders driver convenience. Moreover, the risk of an overfitting problem can when a large number of features are included, and thus, we performed an exhaustive feature selection by limiting the number to five \cite{guyon2003introduction}. Interestingly, all of the crafted features were selected by the feature selection algorithm and the remaining two features were kurtosis and spectral brightness. Table \ref{tab:selectedfeature} shows the features selected for \texttt{HODOR} according to the modulation scheme. Kurtosis is a measure of the peakedness of the sampled signal in the time domain. As the signal propagates through the air, noise signals and multipath signals affect signal quality. In addition, since the passband signal and noise signals are also intensified by an analog amplifier, and external amplification affects the kurtosis of the signal. Therefore, single-band relay and amplification attacks inhere greater kurtosis values than a legitimate signal. The kurtosis is calculated as follows.}   
\begin{equation}
Kurtosis =E\left [ \left ( \frac{d_{RMS}-\mu}{\sigma} \right )^{4} \right ]
\end{equation}
where $\mu$ and $\sigma$ are the mean and standard deviation of the $d_{RMS}$, respectively.
Spectral brightness is the amount of spectral energy corresponding to frequencies higher than a given cut-off threshold. In a playback attack, the adversary records the legitimate signal. In this process,the baseband RF signal goes through an analog-to-digital converter (ADC) and is digitally sampled. However, digital sampling introduces a quantization error due to the sampling rate and vertical resolution of ADC. When the attack is mounted, these samples go through a digital-to-analog converter (DAC) and are reconstructed to an analog baseband signal. At this point, the quantization error introduced during digital sampling affects the spectral density of the reconstructed signal \cite{widrow2008quantization}. Spectral brightness is calculated as follows.
\begin{equation}
Spectral Brightness = \sum_{f=f_{th}}^{0.5\times f_{s}}\left | D_{RMS}[f] \right |^{2}
\end{equation}
where $f_{th}$ is the threshold frequency and $f_{s}$ is the sampling frequency. In our evaluation, we assigned $f_{th}$ as $0.1\times f_{s}$.

\begin{algorithm}[!t]
\begin{minipage}{0.48\textwidth}
\caption{\texttt{Attack detection for the PKES system}}\label{Attack detection for PKE}
\label{algorithm}
\begin{algorithmic}[1]
\Function{Semi-supervised learning (S: a set of signals)}{}
%\State $\text{$F_{PKE}$ = [$f_{peak}$, $SNR_{dB}$, $Spec.Brightness$]}$
\For {i=1 to $|S|$}
	\State $\textit{$d_{RMS}$} \gets \text{preprocessing $(s_{i})$ $(s_{i}\in S)$}$
	\State $\text{$N_{PKES}^{i}$}\gets\text{FeatureExtraction $(d_{RMS},F_{PKES})$}$\
	\State \text{/* $F$ : Selected features */}
	\State \text{/* $N$ : Extracted feature set */}
	\EndFor
\State \textbf{end for}
\State $\text{$\mathbb{C}_{PKES}$} \gets \text{Training ($N_{PKES}$)} $
\State $\text{$\mu_{PKES}, \sigma_{PKES}$} \gets \text{NPC ($N_{PKES}$)} $
\State $\text{/* $\mathbb{C}$ : Classifier */}$
\State \Return $\mathbb{C}_{PKES}, \mu_{PKES}, \sigma_{PKES}$\\
\textbf{end function}
\EndFunction
\item[]
\Function{PKES system attack detection ($s$: received signal)}{}
\State $\textit{$d_{RMS}$} \gets \text{preprocessing $(s)$}$
\State $\text{$N_{PKES}$}\gets\text{Feature Extraction $(d_{RMS},F_{PKES})$}$
\State $\text{$O_{PKES}$} \gets \text{$\mathbb{C}_{PKES}$ ($N_{PKES}$)}$
\State $\text{$O_{PKES}$} \gets \frac{\left | O_{PKES}-\mu_{PKES}  \right |}{\sigma_{PKES}}$
\If {\textit{$O_{PKES} > \Gamma_{PKES}$}} \hspace{0.5cm}\text{/* $\Gamma$ : Threshold */}
\State \Return Reject      \hspace{0.5cm}\text{/*Attack*/}
\Else
\State \Return Accept      \hspace{0.4cm}\text{/*No Attack*/}
\EndIf
\State \textbf{end if}\\
\textbf{end function}
\EndFunction
\end{algorithmic}
\end{minipage}
\end{algorithm}

\begin{table}[ht]
%\normalsize
\centering
\caption{Features used for each modulation scheme}
\label{tab:selectedfeature}
\resizebox{0.7\columnwidth}{!}{%
\renewcommand{\arraystretch}{1.2}
\begin{tabular}{@{}ccc@{}}
\toprule
\begin{tabular}[c]{@{}c@{}}Modulation\\ Scheme\end{tabular} & FSK                                                                                           & ASK                                                                                                              \\ \midrule
\begin{tabular}[c]{@{}c@{}}Selected\\ Features\end{tabular} & \begin{tabular}[c]{@{}c@{}}$f_{peak}$\\ Kurtosis\\ Spec. Brightness\\ $SNR_{dB}$\end{tabular} & \begin{tabular}[c]{@{}c@{}}$f_{peak}$\\ Kurtosis\\ $f_{c}^{offset}$\\ Spec. Brightness\\ $SNR_{dB}$\end{tabular} \\ \bottomrule
\end{tabular}
}
\end{table}

\subsection{Training and Attack Detection}
Before attack detection, \texttt{HODOR} requires a one-class classifier, which facilitates attack detection. One-class classifiers are trained with a set of features derived exclusively from a legitimate key fob. Feature extraction during training can only occur via the legitimate key fobs, and as such, the classifiers are created via semi-supervised learning.
Table \ref{tab:selectedfeature} shows the features used for each modulation scheme. After the classifiers are trained, \texttt{HODOR} assigns a threshold for each classifier. Considering implementation in a real vehicle, it is necessary to assign equal thresholds to a specific key fob model. For this requirement, \texttt{HODOR} performs z-normalization on the output of the classifier, to compensate for the difference of feature distribution between the key fobs. Z-normalization calculates a z-score which has a distribution with a mean of 0 and a standard deviation of 1. To set a normalization parameter mean and standard deviation, inspired by the $k$-fold cross validation \cite{kohavi1995study}, \texttt{HODOR} randomly selects 90 percent of the legitimate data set for training and 10 percent of the legitimate data for testing. After repeating 10 times to accumulate the output of the legitimate test data, \texttt{HODOR} calculates the mean ($\mu$) and standard deviation ($\sigma$) of the corresponding key fob. We denote this process as \textit{Normalization Parameter Calculation (NPC)}. In the attack detection phase, output ($x$) of a classifier from a newly received signal is normalized as $\frac{\left | x-\mu \right |}{\sigma }$.
If any output from the classifier is not within the indicated threshold ($\Gamma$), the corresponding input is considered malicious. We set an adequate threshold for each keyless entry system through evaluation, as shown in the following chapter. Finally, \texttt{HODOR} rejects the door unlock request if at least one classifier is deemed malicious. 
Algorithm \ref{algorithm} illustrates \texttt{HODOR} operation during training and attack detection.

%\vspace*{-0.3cm}
%-------------------------------------------------------------------------------
\section{Evaluation}
\label{sec:eval}
%-------------------------------------------------------------------------------
In this section, we report the evaluation results for \texttt{HODOR} to show that the system accurately detects attacks defined in Section \ref{subsec:attack model}. In addition, we performed further evaluations to demonstrate how \texttt{HODOR} handles environmental factors, such as temperature variations, NLoS conditions, and battery aging.

\begin{figure}[t]
\centering
\subfloat[]{%
  \includegraphics[clip,height=3.6cm]{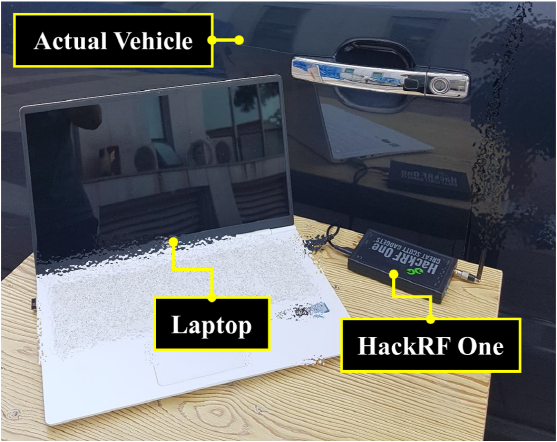}%
  \label{fig:ex_setup_hodor}
}
\subfloat[]{%
  \includegraphics[clip,height=3.6cm]{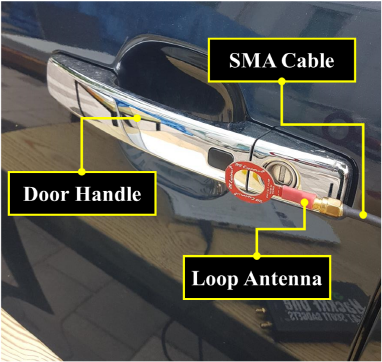}%
  \label{fig:ex_setup_handle}
}
\caption{Experimental setup for single-band relay attack simulation: (a) Signal acquisition setup on the vehicle side, (b) LF-band signal relay using a loop antenna and SMA cable}
\label{fig:ex_setup_single_relay}
\end{figure}

\subsection{Experimental Setup}
\indent\textbf{Keyless Entry System. } We performed a series of experiments on actual vehicles, a 2014 Kia Soul and 2016 Volkswagen Tiguan. Both vehicle models are equipped with a PKES system. In the case of the Soul, an FSK modulation was employed, and a center frequency of 433.92MHz with a frequency deviation of 30kHz was assigned for UHF-band RF communication. In the Tiguan, ASK modulation with a center frequency of 433.92MHz was employed for UHF-band RF communication.\par

\begin{figure}[!t]
  \centering
  \includegraphics[clip,width=\columnwidth]{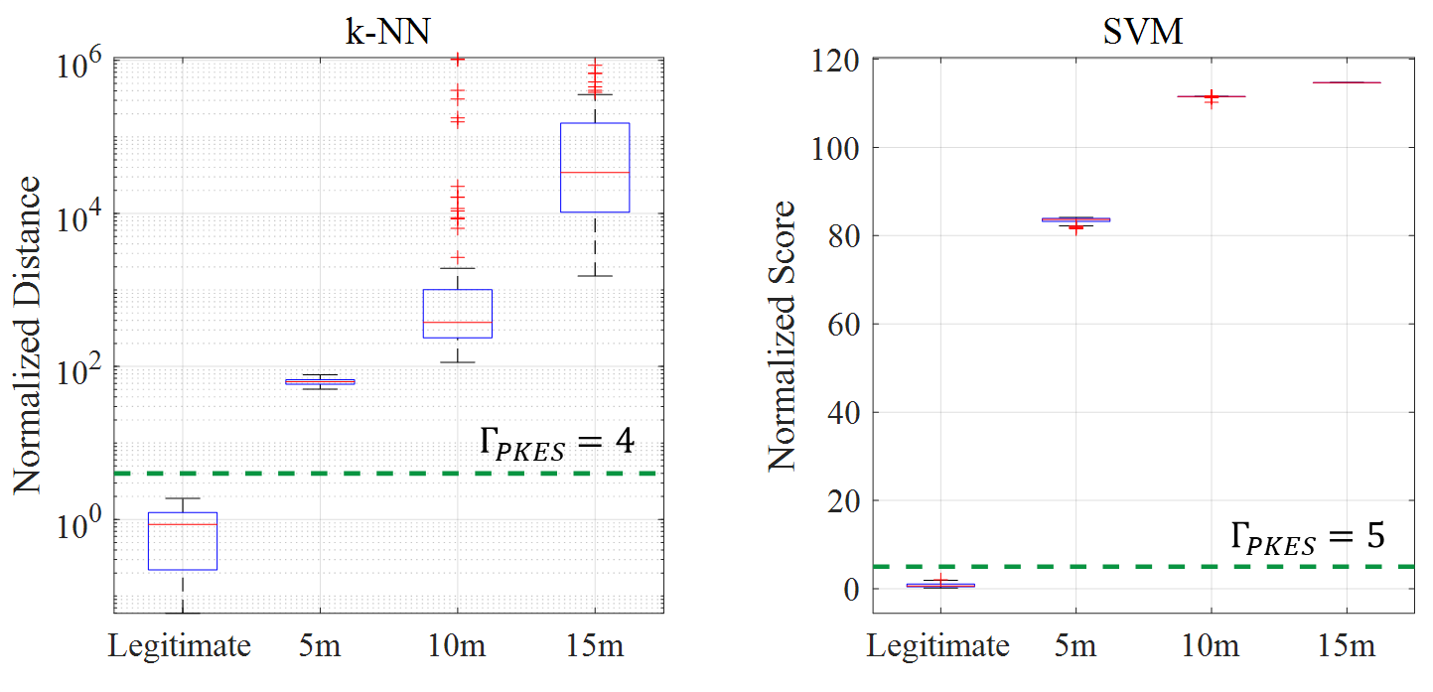}%
  \caption{Output distributions of k-NN and SVM algorithms as a function of distance in a single-band relay attack}
  \label{fig:Single_relay}
\end{figure}

\indent\textbf{{RF Signal Receiver and Transmitter. }} For the evaluation of \texttt{HODOR}, two types of software-defined radio (SDR) devices were used for the transmission and acquisition of the UHF-band RF signals. SDR is a radio communication system that replaces hardware components with a software module. HackRF One \cite{gadgets2017hackrf} was used to sample the UHF-band RF signals, and the other HackRF One coupled with a universal software radio peripheral (USRP) X310 \cite{USRPSoft60:online} was used to generate UHF-band RF signals that were to be simulated as attack signals. With GNU Radio \cite{blossom2004gnu}, the preprocessing phase of \texttt{HODOR} was implemented in virtual hardware components. We set the sample rate of SDR to 5M samples/s in both vehicle models. Since specification of the communication system is different for the two vehicle models, we assigned different parameters for each vehicle. \textcolor{black}{The key fob of the Kia Soul was implemented with a bit rate of 3\si{kbps} and a frequency deviation of 30\si{\kHz} using FSK modulation. As a result, the frequency range of the baseband signal begins at 27\si{\kHz} to 33\si{\kHz}. Ideally, the transition width of a filter should be zero. However, since the practical communication system has an inevitable frequency offset, the receiver must be designed with a wider bandwidth than in ideal scenarios. Accordingly, we set the bandwdith of the bandpass filter to have greater margins and found the specific parameter heuristically. Therefore, we set the high, low cut-off frequencies and transition width of the bandpass filter to 15\si{\kHz}, 45\si{\kHz}, and 10\si{\kHz} respectively. On the other hand, the key fob of the Volkswagen Tiguan uses a MEGAMOS ID 48 transceiver \cite{verdult2013dismantling} which utilizes ASK modulation with a bit rate of 3.5kbps. Therefore, the baseband signal of a Tiguan has a 7\si{kHz} bandwidth. As with the KIA Soul, considering the frequency offset that occurs in a practical system, we set the cut-off frequency of the lowpass filter as 20\si{kHz}.} In addition, the LF-band RF signals were relayed by an SMA cable \cite{SMACable54:online} and a loop antenna \cite{loopantenna:online} to simulate the relay attack. Finally, three RF amplifiers were used to simulate an amplification attack, in which the communication range of a key fob was extended.
In our experimental setup, \textit{we confirmed that the vehicle verifies an attack signal as legitimate in every trial.}
%Fig. \ref{fig:Experimental_setup} shows the experimental setup to transmit, receive, and extend RF signals.\par

\indent\textbf{Classification Algorithm. } Classification algorithms are usually categorized as one-class or multi-class classification. Since it is impossible to train for all cases of malicious attacks before they occur, the classifiers should be trained with a set of features from a legitimate key fob only. 
In other words, a semi-supervised one-class classification is needed for \texttt{HODOR} to cover unknown attacks.
In our evaluation, a one-class support vector machine (SVM) and k-nearest neighbors (k-NN) algorithms were used \cite{khan2014one}. The SVM and k-NN algorithms were performed with the default parameters provided by MatLab 2017a \cite{Trainbin43:online}. More specifically, a residual basis function (RBF) was used for SVM algorithms, and standardized euclidean distance was applied for k-NN algorithms, whose  parameter $k$ was set to 1. For each classifier, we collected a set of 100 UHF-band RF signals from a legitimate key fob from a one-meter distance, and the classifiers were trained with them. As with the training data set, we collected a set of 100 attack signals in every attack simulation.\par

\begin{figure}[!t]
  \centering
  \includegraphics[clip,width=0.75\columnwidth]{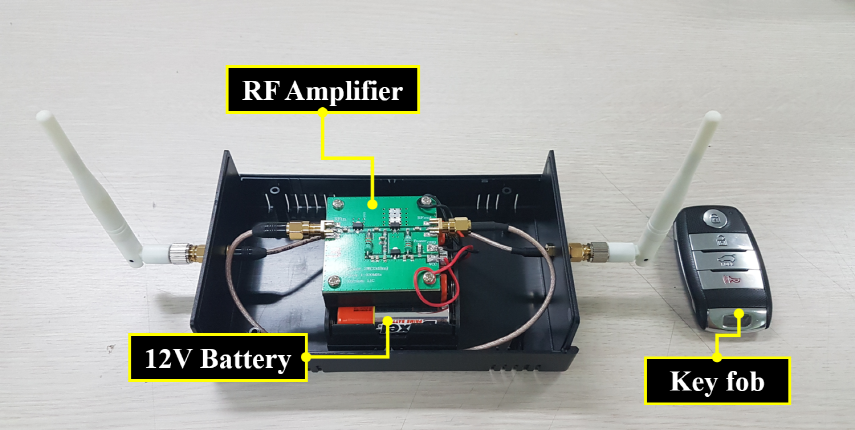}%
  \caption{Experimental setup for amplification attack simulation on the key fob side}
  \label{fig:ex_setup_amp_keyfob}
\end{figure}

\begin{figure}[!t]
  \centering
  \includegraphics[clip,width=0.95\columnwidth]{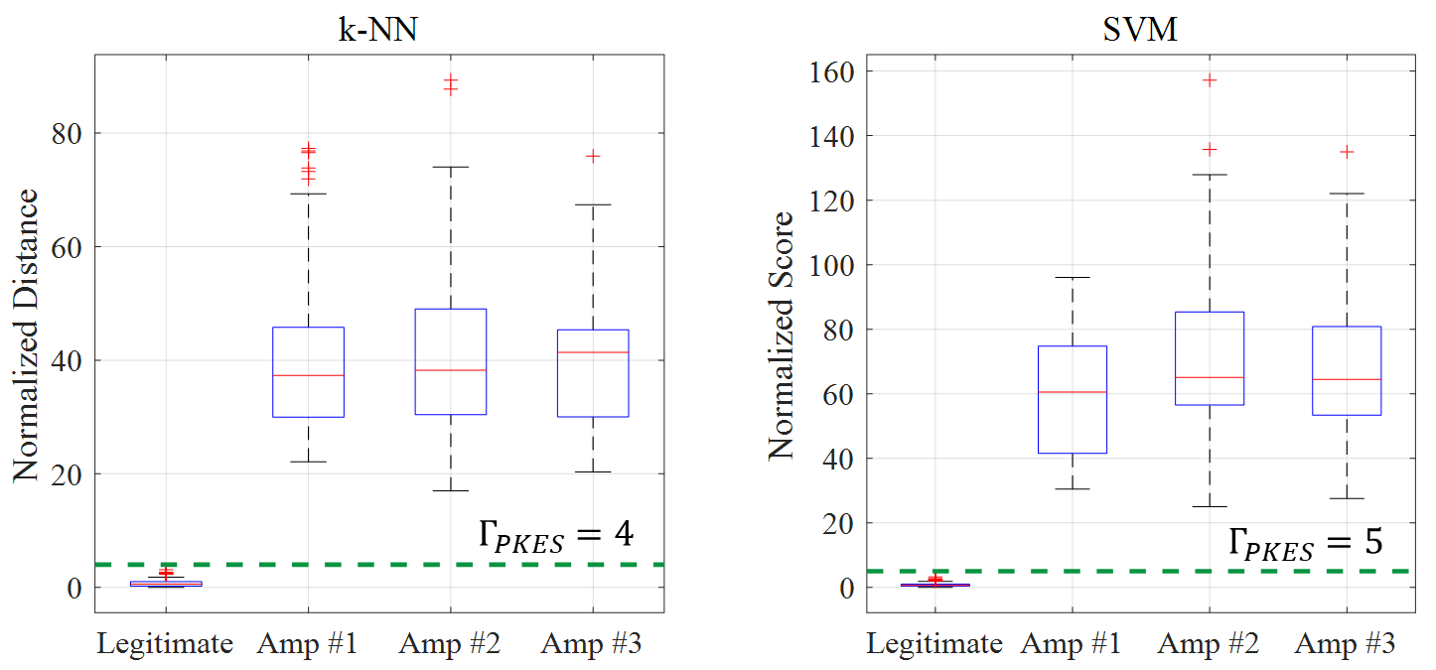}%
  \caption{Output distribution of the k-NN and SVM algorithms as a function of amplifiers in an amplification attack}
  \label{fig:dual_relay_amp}
\end{figure}

\begin{figure}[!t]
  \centering
  \includegraphics[clip,width=0.9\columnwidth]{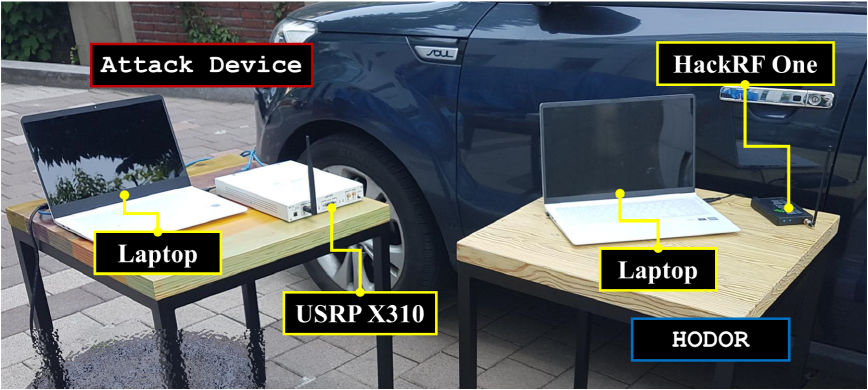}%
  \caption{Experimental setup for digital relay attack simulation}
  \label{fig:ex_digi_relay_setup}
\end{figure}

\indent\textbf{Performance Metric. } Statistical measures of classification test performance were measured by standard metrics, such as true positive rate (TPR), true negative rate (TNR), false positive rate (FPR), and false negative rate (FNR) \cite{guyon2003introduction}. In our evaluation, TP refers to the case in which \texttt{HODOR} identified an attack signal as an attack. Alternatively, TN refers to the case in which \texttt{HODOR} considered a legitimate signal as legitimate. FP refers to the case in which \texttt{HODOR} considered a legitimate signal as an attack and FN refers to the case in which \texttt{HODOR} considered an attack signal as legitimate. It would only take one FN case in the keyless entry system to cause a car theft. Owing to this, we set the objective FNR as 0$\%$, under the belief that FNR should take precedence over FPR.

\subsection{Single-Band Relay Attack Detection}\label{subsec:single-relay}
To simulate the single-band relay attack, we relayed the LF-band signals to trigger a key fob even if it is out of the LF-band communication range. An SMA cable and RF amplifier were used to minimize the path loss of the LF-band signals.
We then sampled the UHF-band RF signals emitted from a key fob, varying the distance between the vehicle and the key fob (5m, 10m, and 15m). Fig. \ref{fig:ex_setup_single_relay} shows the experimental setup for the single-band relay attack simulation. HackRF One is located on the vehicle side and controlled by a laptop. Since the signal attenuation becomes larger as the distance increases, we utilized RF amplifiers to increase LF-band signal strength to relay it to the key fob located more than 10 meters away. It should be noted that the maximum range that the UHF-band signal in the PKES system could transmit was at most 15m in our experimental setup. In addition, we set the distance between HackRF One and the key fob to one meter under LoS conditions for capturing legitimate signals.
Fig. \ref{fig:Single_relay} shows the output distribution of the k-NN and SVM algorithms as a function of distance. Due to the nature of the PKES system, a transmission from an out-of-range key fob is considered an attack on the PKES system. 
For the UHF-band RF signals measured at a distance of 5m, the k-NN and SVM algorithms both output an FPR of 0$\%$, with an FNR of 0$\%$ at thresholds ($\Gamma _{PKES}$) of 4 and 5, respectively.
Furthermore, both algorithms with the same threshold output an FPR of 0$\%$, with an FNR of 0$\%$ where the UHF-band RF signals were captured at distances of 10 or 15m. As the distance increased, it became easier to detect a single-band relay attack. Thus, we conclude that \texttt{HODOR} is able to effectively detect single-band relay attacks.

%% 여기부터 할차례_20200107
\subsection{Dual-Band Relay Attack Detection}\label{subsec:dual-relay}
We evaluate attack detection performance against two types of dual-band attacks, (i.e., amplification attack, digital relay attack) which are mentioned in Section \ref{subsec:attack model}.
 
\begin{figure}[t]
  \centering
  \includegraphics[clip,width=1\columnwidth]{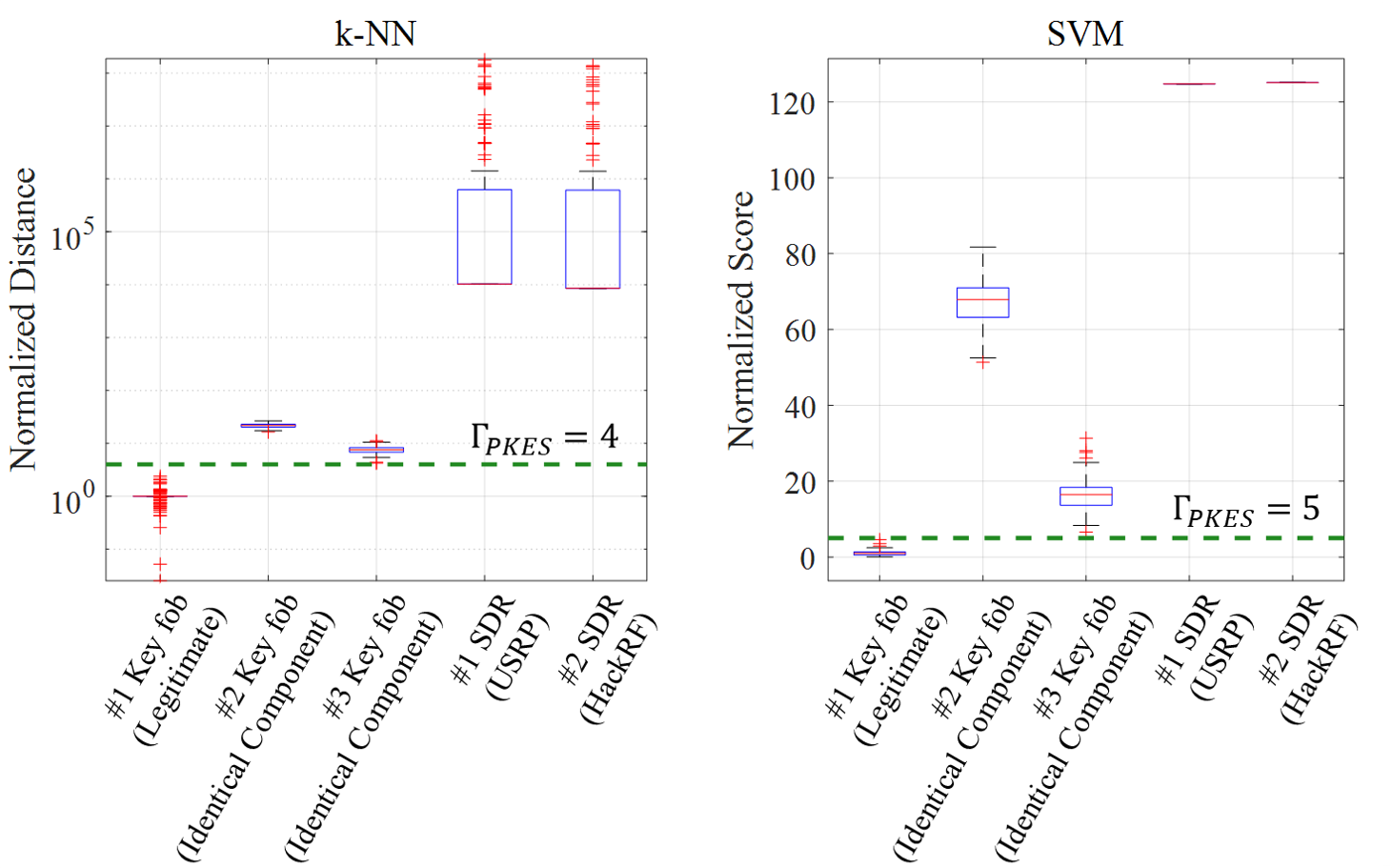}%
  \caption{Output distribution of the k-NN and SVM algorithms as a function of devices in a digital relay attack}
  \label{fig:dual_relay}
\end{figure}

\subsubsection{Amplification attacks}
Since a single-band relay attack is only possible within the communication range of a key fob, a victim might easily become suspicious of foul play. To avoid alerting their victims, adversaries are more likely to adapt a dual-band attack strategy. A dual-band relay attack involves relaying the UHF-band signals of a key fob as well as the LF-band signals of a vehicle. Even though a key fob is placed out of the communication range, a dual-band relay attack can still successfully unlock doors. In our experiments, we first applied RF amplifiers to extend the UHF-band communication range. In an amplification attack scenario, the adversary amplifies and forwards the UHF-band signals to the vehicle. To minimize the path loss between the key fob and RF amplifier, we directly place the RF amplifier next to the key fob. Therefore, the UHF-band RF signals were transmitted much longer distances than the original communication range permitted. Fig. \ref{fig:ex_setup_amp_keyfob} shows the relay module for the amplification attack simulation. The 12V battery pack supplies DC voltage to the amplifier, and the experimental setup on the vehicle side is identical Fig. \ref{fig:ex_setup_single_relay}. Deviating slightly from the attack model in Section \ref{subsec:attack model}, for experimental convenience, we relayed the LF-band signal using an SMA cable. Since \texttt{HODOR} only analyzes the UHF-band signal, this experimental setup is equivalent to an amplification attack model. We employed several Low Noise Amplifiers (LNA) on the UHF band in a commercial market. Each amplifier (Amp $\#$1, Amp $\#$2, Amp $\#$3) used for the attack simulation had 30dB, 60dB and 64dB gains, respectively \cite{LNA:online, 2wlna:online, lna64dbgain:online}. As mentioned in Section \ref{subsec:attack model}, \texttt{HODOR} sampled the amplified UHF-band signals, which had been directly injected by the adversary on the key fob side. We assumed that the adversary can adjust the SNR level using a high-quality amplifier or directional antenna, and set the distance between the adversary and vehicle to the point where the SNR level is equal to that of legitimate signals. This implies that when the adversary injects the attack signal with a higher SNR than the legitimate signal, \texttt{HODOR} can easily detect an amplification attack. We found that the adversary can achieve the same SNR level as the legitimate signal at a distance of between 20 and 25 meters. Then, we sampled the forwarded signals on the vehicle side. Fig. \ref{fig:dual_relay_amp} shows the output distributions of the k-NN and SVM algorithms as a function of the amplifiers. As seen in Fig. \ref{fig:dual_relay_amp}, even if the distance between the key fob and vehicle is larger than the maximum distance of a single-band relay attack, the normalized output distance/score is much closer to the legitimate case. Nevertheless, the k-NN and SVM algorithms both output an FPR of 0\% and FNR of 0\% at thresholds ($\Gamma _{PKES}$) of 4 and 5, respectively. Therefore, \texttt{HODOR} is still able to effectively detect amplification attacks.

\begin{table}[!t]
\centering
\caption{Output distribution of a digital relay attack against the PKES system of the Kia Soul}
\label{tab:digital_relay_soul}
\resizebox{1\columnwidth}{!}{%
\renewcommand{\arraystretch}{1.2}
\begin{tabular}{@{}cccccccc@{}}
\toprule
\multirow{2}{*}{Algorithm} & \multirow{2}{*}{\begin{tabular}[c]{@{}c@{}}Legitimate\\ Device\end{tabular}} & \multicolumn{5}{c}{Output Distribution (mean / std)}                                                                                                                                                                                                                                                            & \multirow{2}{*}{\begin{tabular}[c]{@{}c@{}}FPR\\ (\%)\end{tabular}} \\ \cmidrule(lr){3-7}
                           &                                                                              & \#1 Key fob                                           & \#2 Key fob                                                   & \#3 Key fob                                                   & \#1 SDR                                                    & \#2 SDR                                                    &                                                                     \\ \midrule
\multirow{2}{*}{k-NN}      & \#2 Key fob                                                                  & \begin{tabular}[c]{@{}c@{}}22.29\\ /1.7\end{tabular}  & \textbf{\begin{tabular}[c]{@{}c@{}}0.97\\ /0.63\end{tabular}} & \begin{tabular}[c]{@{}c@{}}11.72\\ /1.76\end{tabular}         & \begin{tabular}[c]{@{}c@{}}2.32e+7\\ /1.15e+7\end{tabular} & \begin{tabular}[c]{@{}c@{}}2.42e+7\\ /1.15e+7\end{tabular} & 0.19                                                                \\ \cmidrule(l){2-8} 
                           & \#3 Key fob                                                                  & \begin{tabular}[c]{@{}c@{}}8.89\\ /1.49\end{tabular}  & \begin{tabular}[c]{@{}c@{}}12.31\\ /1.96\end{tabular}         & \textbf{\begin{tabular}[c]{@{}c@{}}0.91\\ /0.49\end{tabular}} & \begin{tabular}[c]{@{}c@{}}1.21e+7\\ /6.77e+6\end{tabular} & \begin{tabular}[c]{@{}c@{}}1.61e+7\\ /6.06e+6\end{tabular} & 0.29                                                                \\ \midrule
\multirow{2}{*}{SVM}       & \#2 Key fob                                                                  & \begin{tabular}[c]{@{}c@{}}81.16\\ /6.47\end{tabular} & \textbf{\begin{tabular}[c]{@{}c@{}}0.91\\ /0.9\end{tabular}}  & \begin{tabular}[c]{@{}c@{}}33.32\\ /5.68\end{tabular}         & \begin{tabular}[c]{@{}c@{}}153.12\\ /2.28e-13\end{tabular} & \begin{tabular}[c]{@{}c@{}}158.03\\ /1.99e-13\end{tabular} & 3.23                                                                \\ \cmidrule(l){2-8} 
                           & \#3 Key fob                                                                  & \begin{tabular}[c]{@{}c@{}}15.56\\ /3.57\end{tabular} & \begin{tabular}[c]{@{}c@{}}24.48\\ /4.92\end{tabular}         & \textbf{\begin{tabular}[c]{@{}c@{}}0.6\\ /0.55\end{tabular}}  & \begin{tabular}[c]{@{}c@{}}94.91\\ /1.14e-13\end{tabular}  & \begin{tabular}[c]{@{}c@{}}96.73\\ /8.56e-14\end{tabular}  & 1.17                                                                \\ \bottomrule
\end{tabular}
}
\end{table}

\begin{table}[]
\centering
\caption{Output distribution of a digital relay attack against the PKES system of the Volkswagen Tiguan}
\label{tab:dual_relay_pke_Tiguan}
\resizebox{1\columnwidth}{!}{%
\renewcommand{\arraystretch}{1}
\begin{tabular}{@{}ccccccc@{}}
\toprule
\multirow{2}{*}{Algorithm} & \multirow{2}{*}{\begin{tabular}[c]{@{}c@{}}Legitimate\\ Device\end{tabular}} & \multicolumn{4}{c}{Output Distribution (mean / std)}                                                                                                                                                                                                               & \multirow{2}{*}{\begin{tabular}[c]{@{}c@{}}FPR\\ (\%)\end{tabular}} \\ \cmidrule(lr){3-6}
                           &                                                                              & \#1 Key fob                                                     & \#2 Key fob                                                     & \#1 SDR                                                 & \#2 SDR                                                 &                                                                     \\ \midrule
\multirow{2}{*}{k-NN}      & \#1 Key fob                                                                  & \textbf{\begin{tabular}[c]{@{}c@{}}18.24\\ /12.77\end{tabular}} & \begin{tabular}[c]{@{}c@{}}123.22\\ /2.77\end{tabular}          & \begin{tabular}[c]{@{}c@{}}130.03\\ /4.34\end{tabular}  & \begin{tabular}[c]{@{}c@{}}121.8\\ /3.14\end{tabular}   & 0                                                                   \\ \cmidrule(l){2-7} 
                           & \#2 Key fob                                                                  & \begin{tabular}[c]{@{}c@{}}140.74\\ /3.83\end{tabular}          & \textbf{\begin{tabular}[c]{@{}c@{}}11.72\\ /15.72\end{tabular}} & \begin{tabular}[c]{@{}c@{}}112.97\\ /13.61\end{tabular} & \begin{tabular}[c]{@{}c@{}}112.97\\ /13.62\end{tabular} & 0                                                                   \\ \midrule
\multirow{2}{*}{SVM}       & \#1 Key fob                                                                  & \textbf{\begin{tabular}[c]{@{}c@{}}21.52\\ /16.62\end{tabular}} & \begin{tabular}[c]{@{}c@{}}133.24\\ /0.49\end{tabular}          & \begin{tabular}[c]{@{}c@{}}134.72\\ /0.47\end{tabular}  & \begin{tabular}[c]{@{}c@{}}133.6\\ /0.58\end{tabular}   & 0                                                                   \\ \cmidrule(l){2-7} 
                           & \#2 Key fob                                                                  & \begin{tabular}[c]{@{}c@{}}119.58\\ /0.02\end{tabular}          & \textbf{\begin{tabular}[c]{@{}c@{}}20.07\\ /27.76\end{tabular}} & \begin{tabular}[c]{@{}c@{}}118.74\\ /0.46\end{tabular}  & \begin{tabular}[c]{@{}c@{}}118.74\\ /0.46\end{tabular}  & 0                                                                   \\ \bottomrule
\end{tabular}
}
\end{table}

\subsubsection{Digital relay attacks}\label{subsubsec:dual_relay}
To simulate a digital relay attack, we extracted binary information from the ACK signal in Fig. \ref{fig:MF_PKE}. Then, attack signals were injected according to the modulation scheme of the target PKES system using SDR devices. The ACK signal of each key fob contains unique, but static binary information. When the vehicle receives the ACK signal, a number of ECUs are activated to transmit the CAN packets. This \textit{standby} function is implemented in modern vehicles for enhanced driver convenience \cite{cho2018killed}. Based on this observation, and by checking the in-vehicle network (i.e., CAN bus), we confirmed that the vehicle accepts the attack signal. In addition to the SDR devices, we further extended the capability of the digital relay adversary. The strongest adversary would, in theory, be one with access to identical electronic components as the target key fob. In practice, though, the assumption that the digital relay adversary would have the exact same electronic components might be perceived as overly cautious. However, we also evaluated \texttt{HODOR} against this well-equipped, but highly unlikely, hypothetical super adversary. For the Kia Soul, one key fob out of three used in the experiment was chosen as legitimate. The remaining two key fobs and two SDR devices were used to simulate malicious UHF-band RF packets. For example, if the $\#$1 key fob were to be chosen as legitimate, features from the other key fobs would be assumed as an attack. For the Volkswagen Tiguan, one key fob out of the two was chosen as legitimate and the remaining key fob and two SDR devices were used to simulate an attack. Fig. \ref{fig:ex_digi_relay_setup} shows the experimental setup for a digital relay attack simulation. HackRF One was used for signal acquisition, and the USRP and another HackRF One was used for signal injection. All of these SDR devices were controlled by a laptop.
The UHF-band RF signals corresponding to the packets were then sampled and analyzed by \texttt{HODOR}. Fig. \ref{fig:dual_relay} shows the output distributions of k-NN and SVM algorithms as a function of devices used in the simulation of digital relay attacks on the Kia Soul, when the $\#$1 key fob was used as a training set.
As shown in Fig. \ref{fig:dual_relay}, Table \ref{tab:digital_relay_soul}, and Table \ref{tab:dual_relay_pke_Tiguan}, the output from the remaining remote key fobs is closer to that of a legitimate one than to the output from the SDR devices. Especially, features from the $\#$3 key fob of the Soul are closer to that of the $\#$1 key fob than to other devices. This is because the two key fobs were manufactured in the same year and month.
In the case of the Soul, the k-NN and SVM algorithms output produced an average FPR of 0.65$\%$ and an average FPR of 0.27$\%$ with an FNR of 0\% at thresholds ($\Gamma _{PKES}$) of 4 and 5, respectively. In the case of the Tiguan, the k-NN and SVM algorithms both output an average FPR of 0$\%$ with an FNR of 0$\%$ at a threshold ($\Gamma _{PKES}$) of 70 for both. In addition, as mentioned in Section \ref{subsec:attack model}, a cryptographic attack can also be simulated in the same way. Therefore, we also verified that \texttt{HODOR} can effectively detect a cryptographic attack against a PKES system.
As a result, \texttt{HODOR} successfully filtered legitimate and malicious requests from both the amplified and replayed messages. Accordingly, we conclude that \texttt{HODOR} is able to effectively detect dual-band relay attacks and cryptographic attacks.

\begin{figure}[!t]
  \centering
  \includegraphics[clip,width=0.9\columnwidth]{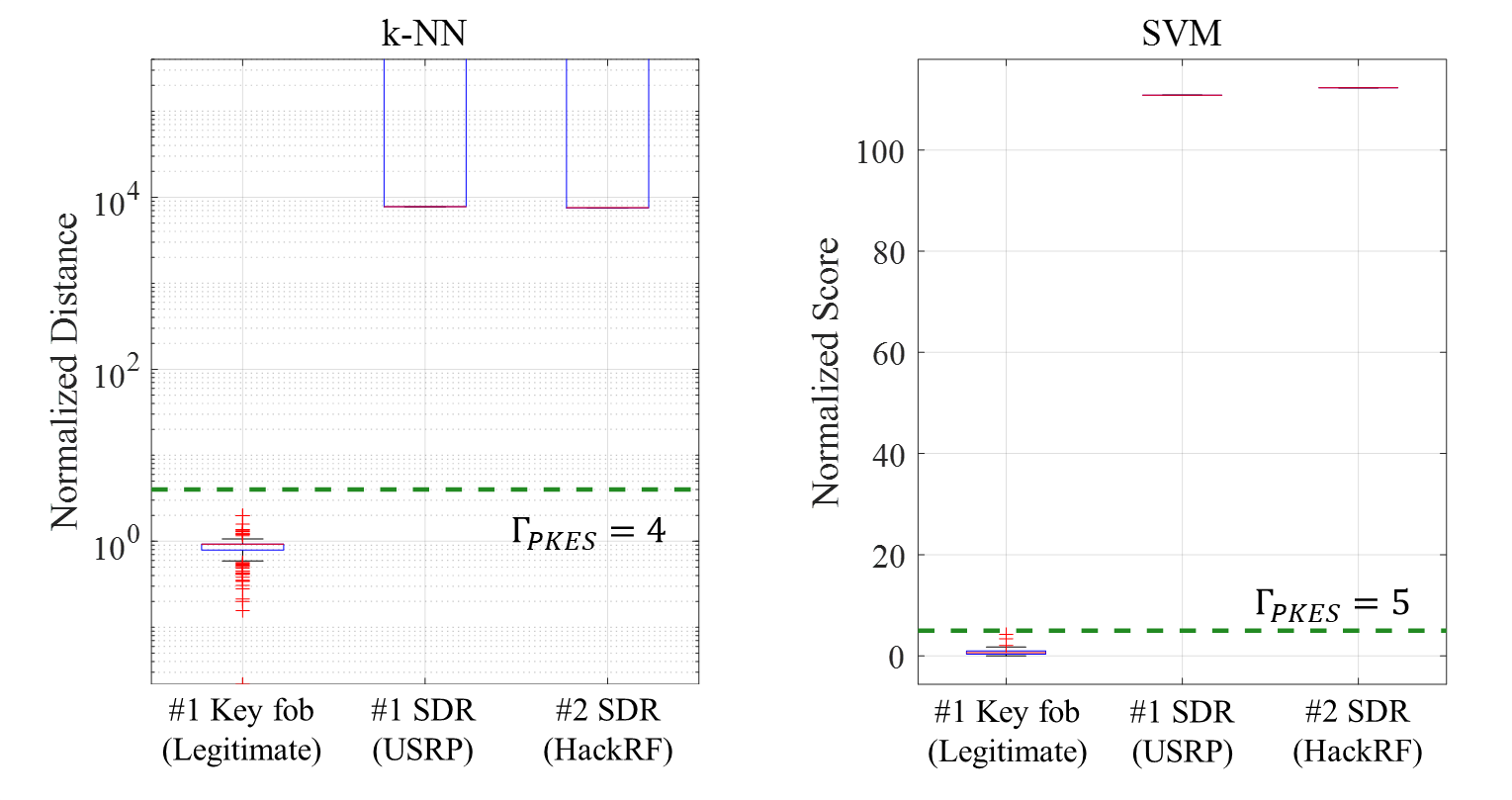}%
  \caption{Output distribution of the k-NN and SVM algorithms as a function of SDR devices in an playback attack on a PKES system}
  \label{fig:playback_pke}
\end{figure}
%% 여기까지 0111
\subsection{Playback Attack Detection}\label{subsec:playbackattack}
During an advanced attack, it is possible that an adversary might attempt to playback the sampled signals whose features most closely resemble the target key fob. This type of attack is a potential threat in PKES systems. For example, an the attack could be mounted when an adversary predicted the next challenge messages through an analysis of several previous challenge messages \cite{alrabady2005analysis}. The adversary can inject such a predicted challenge into the key fob and record the response signals. Then, the adversary goes back to the vehicle and plays back the valid response signals and unlocks the door. SDR devices (i.e., HackRF One and USRP) are employed to transmit the sampled signals. Fig. \ref{fig:playback_pke} shows the output distributions of k-NN and SVM algorithms as a function of SDR devices used in the simulation of playback attacks on the Soul, when the $\#$1 key fob was used as a training set. Table \ref{tab:playback_pke} shows the mean and standard deviation of output distribution according to the vehicle and key fob. The k-NN and SVM algorithms output an average FPR of 0\% and average FPR of 0.35\% with an FNR of 0\% at thresholds of 4 and 5, respectively, in the Soul. In the case of the Tiguan, the k-NN and SVM algorithms both output an average FPR of 0\% with an FNR of 0\% at the threshold of 70. For the SVM algorithm, it is clear that the average output distribution of the $\#1$ SDR device and $\#2$ SDR is closer to the legitimate device than to that of a digital relay attack. From these results, we understand that when the transmission is given a sampled signal (i.e., playback attack) it achieves a closer output distribution to the legitimate key fob than to when the transmission is given a binary code (i.e., digital relay attack). Nevertheless, \texttt{HODOR} still properly detects playback attacks.

\begin{table}[!t]
\centering
\caption{Output distribution of a playback attack against the PKES system of each vehicle}
\label{tab:playback_pke}
\resizebox{1\columnwidth}{!}{%
\renewcommand{\arraystretch}{1}
\begin{tabular}{@{}ccccccc@{}}
\toprule
\multirow{2}{*}{Vehicle} & \multirow{2}{*}{Algorithm} & \multirow{2}{*}{\begin{tabular}[c]{@{}c@{}}Legitimate\\ Device\end{tabular}} & \multicolumn{3}{c}{Output Distribution (mean / std)}                                                                                                                                      & \multirow{2}{*}{\begin{tabular}[c]{@{}c@{}}FPR\\ (\%)\end{tabular}} \\ \cmidrule(lr){4-6}
                         &                            &                                                                              & Legitimate Device                                               & \#1 SDR                                                    & \#2 SDR                                                    &                                                                     \\ \midrule
\multirow{4}{*}{Soul}    & \multirow{2}{*}{k-NN}      & \#2 Key fob                                                                  & \textbf{\begin{tabular}[c]{@{}c@{}}0.99\\ /0.64\end{tabular}}   & \begin{tabular}[c]{@{}c@{}}1.67e+7\\ /5.97e+7\end{tabular} & \begin{tabular}[c]{@{}c@{}}1.38e+7\\ /4.34e+7\end{tabular} & 0                                                                   \\ \cmidrule(l){3-7} 
                         &                            & \#3 Key fob                                                                  & \textbf{\begin{tabular}[c]{@{}c@{}}1.08\\ /0.67\end{tabular}}   & \begin{tabular}[c]{@{}c@{}}1.41e+7\\ /3.43e+7\end{tabular} & \begin{tabular}[c]{@{}c@{}}1.12e+7\\ /3.34e+7\end{tabular} & 0                                                                   \\ \cmidrule(l){2-7} 
                         & \multirow{2}{*}{SVM}       & \#2 Key fob                                                                  & \textbf{\begin{tabular}[c]{@{}c@{}}0.66\\ /0.72\end{tabular}}   & \begin{tabular}[c]{@{}c@{}}86.93\\ /8.56e-14\end{tabular}  & \begin{tabular}[c]{@{}c@{}}85.68\\ /3.28e-15\end{tabular}  & 0.7                                                                 \\ \cmidrule(l){3-7} 
                         &                            & \#3 Key fob                                                                  & \textbf{\begin{tabular}[c]{@{}c@{}}0.7\\ /0.61\end{tabular}}    & \begin{tabular}[c]{@{}c@{}}97.37\\ /8.56e-14\end{tabular}  & \begin{tabular}[c]{@{}c@{}}98.09\\ /1.42e-13\end{tabular}  & 0.7                                                                 \\ \midrule
\multirow{4}{*}{Tiguan}  & \multirow{2}{*}{k-NN}      & \#1 Key fob                                                                  & \textbf{\begin{tabular}[c]{@{}c@{}}16.46\\ /11.4\end{tabular}}  & \begin{tabular}[c]{@{}c@{}}173.37\\ /5.99\end{tabular}     & \begin{tabular}[c]{@{}c@{}}164.79\\ /5.4\end{tabular}      & 0                                                                   \\ \cmidrule(l){3-7} 
                         &                            & \#2 Key fob                                                                  & \textbf{\begin{tabular}[c]{@{}c@{}}12.92\\ /16.71\end{tabular}} & \begin{tabular}[c]{@{}c@{}}83.27\\ /9.99\end{tabular}      & \begin{tabular}[c]{@{}c@{}}83.27\\ /9.99\end{tabular}      & 0                                                                   \\ \cmidrule(l){2-7} 
                         & \multirow{2}{*}{SVM}       & \#1 Key fob                                                                  & \textbf{\begin{tabular}[c]{@{}c@{}}25.43\\ /19.03\end{tabular}} & \begin{tabular}[c]{@{}c@{}}139.66\\ /1e-3\end{tabular}     & \begin{tabular}[c]{@{}c@{}}139.66\\ /3.1e-3\end{tabular}   & 0                                                                   \\ \cmidrule(l){3-7} 
                         &                            & \#2 Key fob                                                                  & \textbf{\begin{tabular}[c]{@{}c@{}}18.79\\ /16.43\end{tabular}} & \begin{tabular}[c]{@{}c@{}}74.6\\ /0.31\end{tabular}       & \begin{tabular}[c]{@{}c@{}}74.6\\ /0.31\end{tabular}       & 0                                                                   \\ \bottomrule
\end{tabular}
}
\end{table}

\subsection{Non-Line-of-Sight (NLoS) Conditions}\label{subsec:nlos}
\begin{figure}[!t]
  \centering
  \includegraphics[clip,width=\columnwidth]{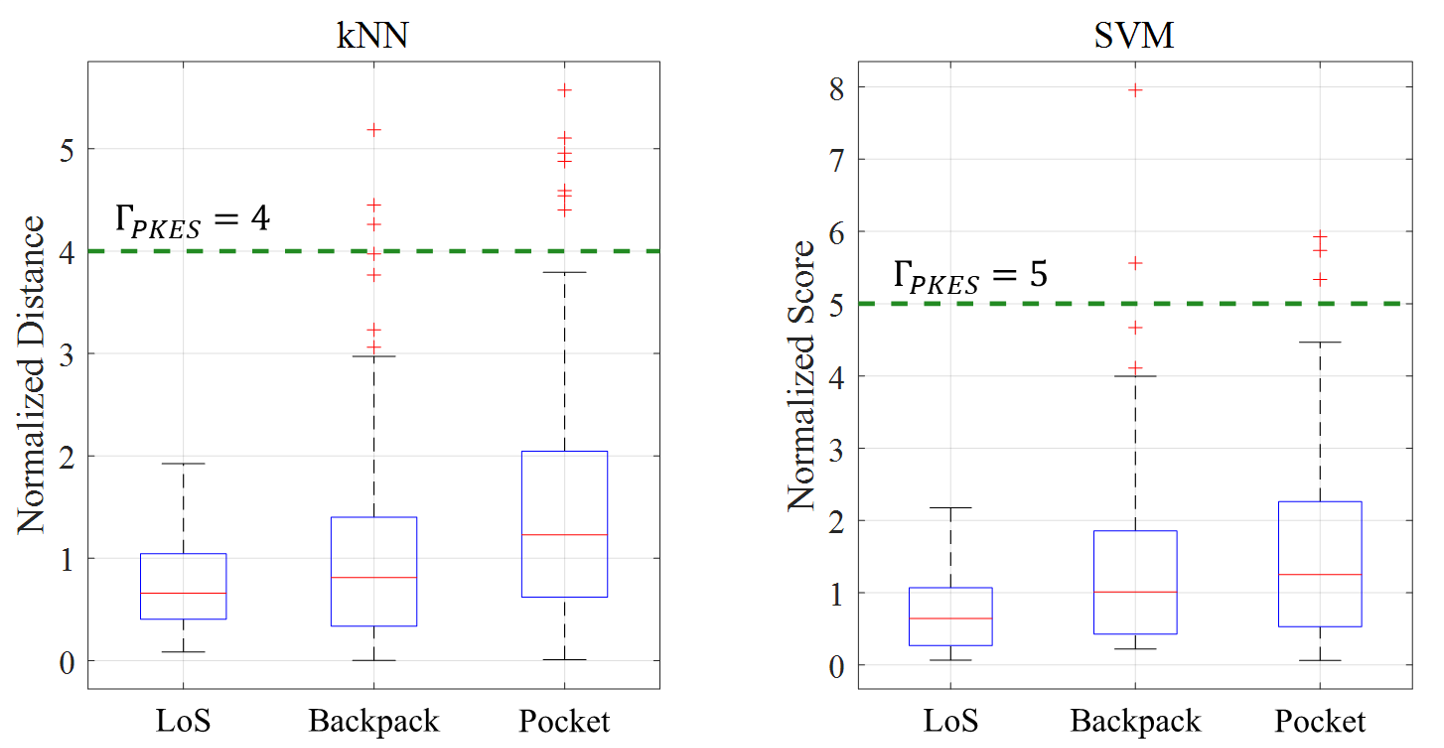}%
  \caption{Output distributions of the k-NN and SVM algorithms as a function of where a key fob is placed}
  \label{fig:nlos_static}
\end{figure}

To show that the features used for the PKES system are robust under an NLoS condition, we sampled UHF-band RF signals from the key fobs placed in a pocket or backpack.
In the PKES system, a car owner is able to unlock doors without physically producing the key fob from its storage location.
The classifier was trained with the UHF-band RF signals sampled under an LoS condition.
Fig. \ref{fig:nlos_static} shows the output distributions of the k-NN and SVM algorithms as a function of where the key fob is placed.
When the key fob is placed in a backpack, the k-NN and SVM algorithms output an FPR of 1.32$\%$ and 1.35$\%$ with an FNR of 0\%. When the key fob is placed in a pocket, the k-NN and SVM algorithms output an FPR of 1.71$\%$ and 1.67$\%$ with an FNR of 0\%. Like in previous experiments, thresholds ($\Gamma _{PKES}$) were respectively assigned to 4 and 5 for each algorithm.
From these results, we conclude that \texttt{HODOR} properly identifies a legitimate door unlock request even when in an NLoS condition. %%

%\vspace*{-0.3cm}
\subsection{Attack Detection in RKE Systems}\label{subsec:rke_detection}
We further analyzed the UHF-band attack signal against the RKE system of the Soul. Cryptographic attacks on RKE systems have been extensively studied in the past decade \cite{garcia2016lock, verdult2012gone, immler2012breaking, benadjila2017one, hicks2018dismantling}. An adversary eavesdrops on several valid packets and exploits the weaknesses of the cryptographic algorithm to extract the secret key. It is also possible for an adversary to attempt to playback the sampled UHF-band signal. One example of a playback attack on an RKE system is a roll-jam attack \cite{ThisHack20:online}. In this attack, the adversary interferes (i.e., jams) some frequencies and plays back valid encrypted messages.\par
To simulate a cryptographic attack, one remote key fob out of three used in the experiment was chosen as legitimate. The remaining two remote key fobs and two SDR devices were used to simulate malicious UHF-band RF packets. In the case of a playback attack, two SDR devices were employed to simulate attack signals. Moreover, we tested the UHF-band RF signals that were sampled at different distances from where they had been sampled during training. Even though the classifier was trained with a training set measured at a distance of one meter, the legitimate remote key fob must still be identified correctly within the communication range of the RKE system. Therefore, we measured the UHF-band signals from the legitimate key fob at a distance of 40 meters from the RF signal receiver under LoS conditions and used this for the test. Since $SNR_{dB}$ and $Kurtosis$ are highly correlated to the distance between the key fob and receiver, they were excluded in this experiment. In other words, a classifier using a different feature set (i.e., $f_{peak}$ and $Spectral Brightness$) was employed for the training. We denote the RKE system classifier as $\mathbb{C}_{RKE}$. As we mentioned in Section \ref{Subsection:Keyless_entry_system}, an identical packet was transmitted from a smart key multiple times for reliable communication. Accordingly, \texttt{HODOR} verifies a newly received signal as an attack when the number of detected preambles exceeds $\left \lfloor \frac{N}{2} \right \rfloor$ among the $N$ received preambles. Therefore, the maximum value of $N$ in our experiment is five. In addition, we set the threshold($\Gamma_{RKE}$) of k-NN and SVM algorithms to 4.5 and 5, respectively. Tables \ref{tab:crypto_rke} and \ref{tab:playback_rke} show the mean and standard deviation of output distribution for both the cryptographic and playback attacks. Even though \texttt{HODOR} did not yield an FPR of 0\% with an FNR of 0\% on the RKE system during experimentation, we conclude that the achieved results are acceptable, as \texttt{HODOR} is designed to act as a support to an existing authentication method in a keyless entry systems.

\begin{table}[!t]
\centering
\caption{Output distribution of a cryptographic attack against the RKE system of the Soul}
\label{tab:crypto_rke}
\resizebox{\columnwidth}{!}{%
\renewcommand{\arraystretch}{1.3}
\begin{tabular}{@{}cccccccc@{}}
\toprule
\multirow{2}{*}{Algorithm} & \multirow{2}{*}{\begin{tabular}[c]{@{}c@{}}Legitimate\\ Device\end{tabular}} & \multicolumn{5}{c}{Output Distribution (mean / std)}                                                                                                                                                                                                                                                                    & \multirow{2}{*}{\begin{tabular}[c]{@{}c@{}}FPR\\ (\%)\end{tabular}} \\ \cmidrule(lr){3-7}
                           &                                                                              & \#1 Key fob                                                   & \#2 Key fob                                                   & \#3 Key fob                                                   & \#1 SDR                                                     & \#2 SDR                                                   &                                                                     \\ \midrule
\multirow{3}{*}{k-NN}      & \#1 Key fob                                                                  & \textbf{\begin{tabular}[c]{@{}c@{}}1.44\\ /2.16\end{tabular}} & \begin{tabular}[c]{@{}c@{}}77.97\\ /9.74\end{tabular}         & \begin{tabular}[c]{@{}c@{}}22.09\\ /5.37\end{tabular}         & \begin{tabular}[c]{@{}c@{}}5.27e+3\\ /511.58\end{tabular}   & \begin{tabular}[c]{@{}c@{}}5.21e+3\\ /220.04\end{tabular} & 2.38                                                                \\ \cmidrule(l){2-8} 
                           & \#2 Key fob                                                                  & \begin{tabular}[c]{@{}c@{}}73.91\\ /5.4\end{tabular}          & \textbf{\begin{tabular}[c]{@{}c@{}}2.36\\ /3.72\end{tabular}} & \begin{tabular}[c]{@{}c@{}}39.17\\ 5.09\end{tabular}          & \begin{tabular}[c]{@{}c@{}}6.56e+3\\ /433\end{tabular}      & \begin{tabular}[c]{@{}c@{}}4.53e+3\\ 99.1\end{tabular}    & 3.57                                                                \\ \cmidrule(l){2-8} 
                           & \#3 Key fob                                                                  & \begin{tabular}[c]{@{}c@{}}23.64\\ /6.48\end{tabular}         & \begin{tabular}[c]{@{}c@{}}49.8\\ /11.83\end{tabular}         & \textbf{\begin{tabular}[c]{@{}c@{}}1.3\\ /1.3\end{tabular}}   & \begin{tabular}[c]{@{}c@{}}5.76e+3\\ /846.02\end{tabular}   & \begin{tabular}[c]{@{}c@{}}5.04e+3\\ /185.41\end{tabular} & 1.79                                                                \\ \midrule
\multirow{3}{*}{SVM}       & \#1 Key fob                                                                  & \textbf{\begin{tabular}[c]{@{}c@{}}1.89\\ /5.88\end{tabular}} & \begin{tabular}[c]{@{}c@{}}47.26\\ /0.75\end{tabular}         & \begin{tabular}[c]{@{}c@{}}21.09\\ /4.47\end{tabular}         & \begin{tabular}[c]{@{}c@{}}47.63\\ /7.13e-14\end{tabular}   & \begin{tabular}[c]{@{}c@{}}43.14\\ /5.7e-14\end{tabular}  & 0                                                                   \\ \cmidrule(l){2-8} 
                           & \#2 Key fob                                                                  & \begin{tabular}[c]{@{}c@{}}22.06\\ /1.78\end{tabular}         & \textbf{\begin{tabular}[c]{@{}c@{}}1.53\\ /1.91\end{tabular}} & \begin{tabular}[c]{@{}c@{}}9.55\\ /1.74\end{tabular}          & \begin{tabular}[c]{@{}c@{}}32.99\\ /4.98e-14\end{tabular}   & \begin{tabular}[c]{@{}c@{}}29.19\\ /4.28e-14\end{tabular} & 0                                                                   \\ \cmidrule(l){2-8} 
                           & \#3 Key fob                                                                  & \begin{tabular}[c]{@{}c@{}}18.14\\ /4.26\end{tabular}         & \begin{tabular}[c]{@{}c@{}}29.64\\ /4.63\end{tabular}         & \textbf{\begin{tabular}[c]{@{}c@{}}1.99\\ /6.67\end{tabular}} & \begin{tabular}[c]{@{}c@{}}47.26\\ /$\approx$0\end{tabular} & \begin{tabular}[c]{@{}c@{}}43.02\\ /7.84e-14\end{tabular} & 0.6                                                                 \\ \bottomrule
\end{tabular}
}
\end{table}
%\vspace*{-0.3cm}
% \usepackage{multirow}
\begin{table}[!t]
\centering
\caption{Output distribution of a playback attack against the RKE system of the Soul}
\label{tab:playback_rke}
\resizebox{0.9\columnwidth}{!}{%
\renewcommand{\arraystretch}{1.1}
\begin{tabular}{@{}cccccc@{}}
\toprule
\multirow{2}{*}{Algorithm} & \multirow{2}{*}{\begin{tabular}[c]{@{}c@{}}Legitimate\\ Device\end{tabular}} & \multicolumn{3}{c}{Output Distribution (mean / std)}                                                                                                                                   & \multirow{2}{*}{\begin{tabular}[c]{@{}c@{}}FPR\\ (\%)\end{tabular}} \\ \cmidrule(lr){3-5}
                           &                                                                              & Legitimate Device                                              & \#1 SDR                                                  & \#2 SDR                                                  &                                                                     \\ \midrule
\multirow{3}{*}{k-NN}      & \#1 Key fob                                                                  & \textbf{\begin{tabular}[c]{@{}c@{}}4.96\\ /32.43\end{tabular}} & \begin{tabular}[c]{@{}c@{}}127.65\\ /141.48\end{tabular} & \begin{tabular}[c]{@{}c@{}}187.96\\ /105.16\end{tabular} & 0                                                                   \\ \cmidrule(l){2-6} 
                           & \#2 Key fob                                                                  & \textbf{\begin{tabular}[c]{@{}c@{}}2.84\\ /5.68\end{tabular}}  & \begin{tabular}[c]{@{}c@{}}105.21\\ /27.43\end{tabular}  & \begin{tabular}[c]{@{}c@{}}40.91\\ /14.2\end{tabular}    & 3.57                                                                \\ \cmidrule(l){2-6} 
                           & \#3 Key fob                                                                  & \textbf{\begin{tabular}[c]{@{}c@{}}5.15\\ /28.55\end{tabular}} & \begin{tabular}[c]{@{}c@{}}88.75\\ /98.71\end{tabular}   & \begin{tabular}[c]{@{}c@{}}191.29\\ /174.67\end{tabular} & 1.79                                                                \\ \midrule
\multirow{3}{*}{SVM}       & \#1 Key fob                                                                  & \textbf{\begin{tabular}[c]{@{}c@{}}1.53\\ /4.87\end{tabular}}  & \begin{tabular}[c]{@{}c@{}}38.96\\ /2.28\end{tabular}    & \begin{tabular}[c]{@{}c@{}}45.08\\ /5.93\end{tabular}    & 0                                                                   \\ \cmidrule(l){2-6} 
                           & \#2 Key fob                                                                  & \textbf{\begin{tabular}[c]{@{}c@{}}2.46\\ /4.31\end{tabular}}  & \begin{tabular}[c]{@{}c@{}}42.04\\ /4.3\end{tabular}     & \begin{tabular}[c]{@{}c@{}}19.37\\ /5.32\end{tabular}    & 3.57                                                                \\ \cmidrule(l){2-6} 
                           & \#3 Key fob                                                                  & \textbf{\begin{tabular}[c]{@{}c@{}}1.75\\ /6\end{tabular}}     & \begin{tabular}[c]{@{}c@{}}36.55\\ /4.88\end{tabular}    & \begin{tabular}[c]{@{}c@{}}34.48\\ /11.86\end{tabular}   & 0                                                                   \\ \bottomrule
\end{tabular}
}
\end{table}
%\vspace*{-0.1cm}
\subsection{Effects of Temperature Variation}
\begin{figure}[t!]
\centering
\subfloat[Experimental setup for temperature variation]{%
  \includegraphics[clip,width=0.65\columnwidth]{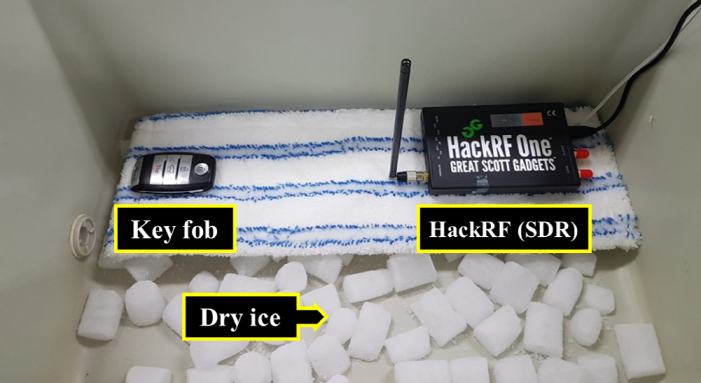}%
  \label{fig:temp_setup}
}

\subfloat[Output distributions of k-NN and SVM algorithms as a function of temperature]{%
  \includegraphics[clip,width=0.99\columnwidth]{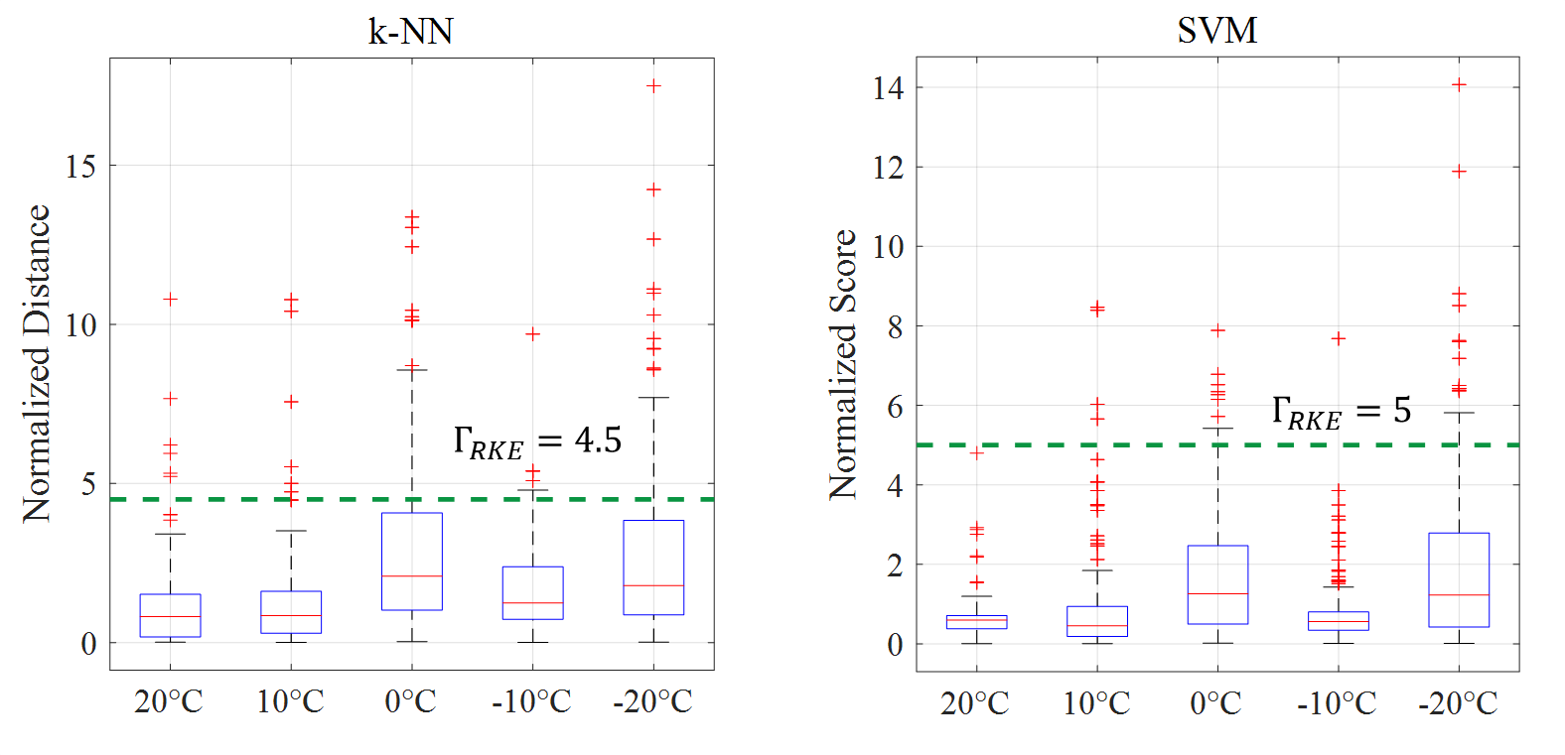}%
  \label{fig:Temp_var_RKE}
}
\label{fig:tmep_rke}
\caption{Experimental results of temperature variation}
\end{figure}

Analog signals are easily affected by external factors. In particular, temperature variation is one of the most critical concerns in device fingerprinting \cite{murdoch2006hot, huang2014blueid, choi2018voltageids}. In order to show that \texttt{HODOR} properly operates under varying temperatures, UHF-band RF signals were sampled in an ice box with dry ice to maintain a certain temperature as shown in Fig. \ref{fig:temp_setup}. 
It should be noted that this setup was only available indoors. Owing to this, evaluation on the PKES system was impossible, as it must occur within the vicinity of a vehicle (i.e., outdoors). 
The UHF-band RF signals for training were measured in degrees, between 20\textdegree{}C to -20\textdegree{}C in intervals of 10\textdegree{}C. 
Fig. \ref{fig:Temp_var_RKE} shows output distributions of the k-NN and SVM algorithms as a function of temperature. 
Unlike an attack detection, UHF-band RF signals measured in terms of varying temperature should be recognized as legitimate. Accordingly, this also means that the output of the RKE system classifier should remain below a given threshold. In Fig. \ref{fig:Temp_var_RKE}, the output distribution at 0\textdegree{}C and -20\textdegree{}C exceeds the threshold. This is not only because of the temperature variation, but also the severe multipath environment caused by the limited space of the ice box. However, as we denoted in Section \ref{subsec:rke_detection}, an RKE packet is considered an attack only if the number of detected preambles exceeds $\left \lfloor \frac{N}{2} \right \rfloor$ among the $N$ received preambles. Therefore, \texttt{HODOR} can achieve moderate FPR under temperature variation.
The k-NN and SVM algorithms output an FPR of 6.36$\%$ and 0.65$\%$ at the same threshold determined in Section \ref{subsec:rke_detection}. Thus, we conclude that the features used in \texttt{HODOR} are operationally robust regardless of temperature variation.

%\vspace*{-0.5cm}
\subsection{Battery Aging}\label{subsec:battery_aging}
\begin{figure}[!t]
  \centering
  \includegraphics[clip,width=\columnwidth]{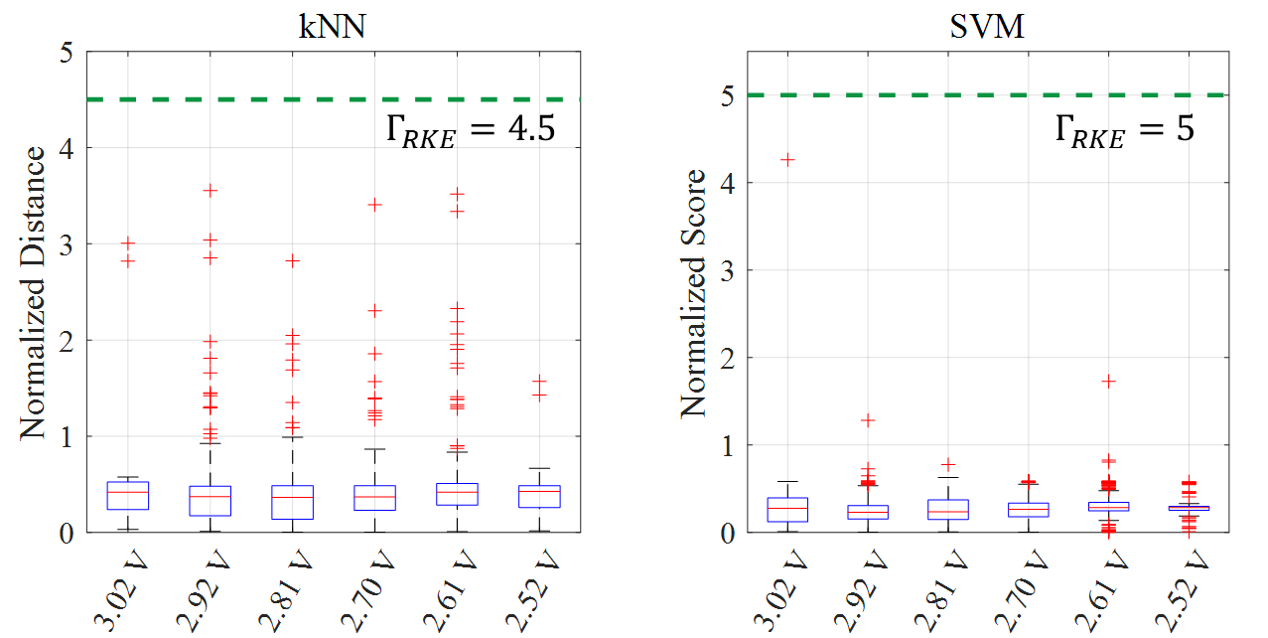}%
  \caption{Output distribution of k-NN and SVM algorithms as a function of battery level}
  \label{fig:batter_level}
\end{figure}

A coin cell battery supplies stable DC voltage to the key fob. However, as the driver uses the key fob, the supplied voltage decreases and may lead to feature variation. For this reason, we further evaluated feature robustness against battery aging. In the case of the Panasonic CR2032 lithium battery, which we selected for our evaluation, the initial voltage was 3V, but decreased to 2.5V over time \cite{Panasoni43:online}. This voltage interval occupies around 97.34$\%$ of the battery duration time. If the voltage drops under 2.5V, the key fob does not operate properly due to insufficient voltage supply. Therefore, using the classifier which had been trained in advance, we tested the UHF-band signals at a specific voltage level between 2.5V and 3V. Fig. \ref{fig:batter_level} shows the output distribution of the k-NN and SVM algorithms as a function of voltage level. Similar to the temperature variation experiment, output of the classifier for the RKE system should be below a given threshold. The k-NN and SVM algorithms both output an FPR of 0$\%$ at the same threshold determined in Section \ref{subsec:rke_detection}. Thus, we conclude that the features used in \texttt{HODOR} are operationally regardless of battery aging.

\begin{table}[!t]
\small
\centering
\caption{Execution time of \texttt{HODOR}}
\label{tab:execution-time}
\resizebox{1\columnwidth}{!}{
\renewcommand{\arraystretch}{1.2}
\begin{tabular}{@{}cccc@{}}
\toprule
\multicolumn{2}{c}{\multirow{2}{*}{Phase}}                                                                          & \multicolumn{2}{c}{Algorithm}           \\ \cmidrule(l){3-4} 
\multicolumn{2}{c}{}                                                                                                & k-NN               & SVM                \\ \midrule
\multirow{5}{*}{\begin{tabular}[c]{@{}c@{}}Feature\\ Extraction\\ (FSK / ASK)\end{tabular}} & $f_{peak}$      & \multicolumn{2}{c}{4$ms$ / 3.85$ms$}       \\
                                                                                            & $f_{c}^{offset}$      & \multicolumn{2}{c}{4$ms$ / 3.55$ms$}    \\
                                                                                            & $SNR_{dB}$            & \multicolumn{2}{c}{130$ms$ / 94$ms$} \\
                                                                                            & $Kurtosis$            & \multicolumn{2}{c}{20$ms$ / 16.2$ms$}  \\
                                                                                            & $Spec. Brightness$ & \multicolumn{2}{c}{5$ms$ / 3.73$ms$}     \\ \midrule
\multirow{2}{*}{\begin{tabular}[c]{@{}c@{}}NPC\\ (FSK / ASK)\end{tabular}}                  & $\mathbb{C}_{PKES}$   & 55$ms$ / 60$ms$  & 43$ms$ / 45.5$ms$  \\
                                                                                            & $\mathbb{C}_{RKE}$    & 50$ms$ / 52$ms$    & 32$ms$ / 34$ms$    \\ \midrule
\multirow{2}{*}{\begin{tabular}[c]{@{}c@{}}Attack Detection\\ (FSK / ASK)\end{tabular}}     & $\mathbb{C}_{PKES}$   & 4.8$ms$ / 4.94$ms$ & .038$ms$ / .04$ms$ \\
                                                                                            & $\mathbb{C}_{RKE}$    & 3.8$ms$ / 4$ms$ & .04$ms$ / .07$ms$ \\ \bottomrule
\end{tabular}
}
\end{table}

\subsection{Execution Time}\label{subsec:executiontime}
\textcolor{black}{We implemented the functions that principally contribute to the total execution time of \texttt{HODOR}. We selected the Raspberry-pi 3B single-board computer \cite{pi2015raspberry} as a reference hardware platform.
This platform is based on a quad core 1.4GHz Cortex-A53 (ARMv8) with 1GB RAM.
Table \ref{tab:execution-time} shows the execution time for each basic function used in \texttt{HODOR}, which were implemented via Python programming. 
The classifiers should be trained and cross-validated in advance. Owing to this, the time for these two processes can be disregarded when it comes to the operation time for verifying a door unlock request.
\begin{figure*}[!t]
  \includegraphics[width=\textwidth]{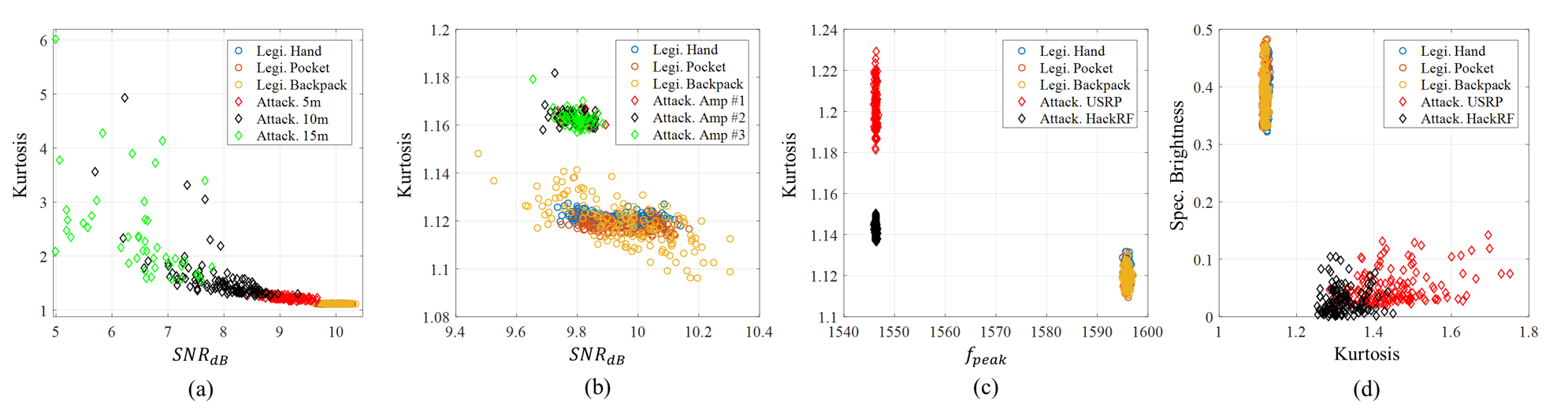}
  \caption{Scatter plot of the top two features of the Soul as a function of an attack scenario: (a) Single-band relay attack, (b) Amplification attack, (c) Digital relay attack, (d) Playback attack}
  \label{fig:scatter}
\end{figure*}
The results show that the k-NN algorithm takes longer to detect attacks. As a result, the total operation time for the verification of the PKES system using FSK modulation is 163.8$ms$ and 159.038$ms$ in the k-NN and SVM algorithms, respectively. These are the sum of feature extraction and attack detection processes. In addition, the RKE system using FSK modulation requires 12.8$ms$ and 9.04$ms$ in the k-NN and SVM algorithms, respectively. In ASK modulation, the total amount of operation time for the verification of the PKES system using ASK modulation is 126.27$ms$ and 121.37$ms$ in the k-NN and SVM algorithms, respectively. In addition, the RKE system using ASK modulation requires 15.13$ms$ and 11.2$ms$ in the k-NN and SVM algorithms, respectively. For the feature extraction phase, similar execution times were obtained because the numbers of samples for both ASK $\&$ FSK modulations were similar to each other. In other words, the duration of the preamble region from which the features are extracted is similar regardless of the modulation schemes. Regarding the \textit{NPC} and \textit{Attack Detection} phases, \texttt{HODOR} has different execution times that correspond to the modulation schemes. In \texttt{HODOR}, the number of used features are different depending on modulation schemes. Because \texttt{HODOR} uses one additional feature, $f_{c}^{offset}$, for ASK-modulated signals, attack detection time for ASK modulation is longer than FSK modulation. However, since the ASK-modulated signal has a much shorter execution time for feature extraction, the total execution time of the ASK-modulated signal is shorter than the FKS-modulated signal. According to \cite{WhyPerce47:online}, humans cannot easily recognize a delay of less than 500$ms$. Thus, it is expected that \texttt{HODOR} would be effective as a support mechanism to existing keyless entry systems without creating a noticeable delay on the user end.
}

\subsection{Feature Importance}\label{subsec:feat_importance}
We minimized the feature set through the exhaustive search in Section \ref{sec:method}. In this subsection, we further evaluated the feature importance as a function of each attack scenario. We employed the Relief algorithm, which is a unique family of filter-style feature-selection algorithms \cite{kira1992feature}. A key idea of the Relief algorithm is to estimate the quality of features according to how well their values distinguish between instances near to each other. Based on the MATLAB implementation of the \texttt{relieff} function, we ranked the features in each attack detection experiment. Table \ref{tab:feature_rank} shows the rankings of the features as a function of each attack scenario. The distribution of the top two features in each attack scenario are represented in Fig. \ref{fig:scatter}. In a single-band relay and amplification attack, $SNR_{dB}$ and kurtosis are effective features to detect an attack. In an amplification attack, even when the adversaries adjust the SNR level to the legitimate signal, \texttt{HODOR} can effectively differentiate the attack signals using the kurtosis feature. In a digital relay attack, $f_{peak}$ has a major role. This is because of the clock difference between the key fob and the SDRs (i.e., USRP and HackRF). Though not as effective as $f_{peak}$, kurtosis is also useful to detect a digital relay attack. In a playback attack, due to the quantization error, spectral brightness and kurtosis are both effective features to differentiate attack signals.

\begin{table}[t]\label{feauter_importance}
\small
\centering
\caption{Feature importance as a function of attack scenario}
\label{tab:feature_rank}
\resizebox{1\columnwidth}{!}{
\def\arraystretch{1.25}
\begin{tabular}{@{}cccccc@{}}
\toprule
\multicolumn{2}{c}{\begin{tabular}[c]{@{}c@{}}Attack\\ Scenario\end{tabular}} & \begin{tabular}[c]{@{}c@{}}Single-band\\ Relay Attack\end{tabular} & \begin{tabular}[c]{@{}c@{}}Amplification\\ Attack\end{tabular} & \begin{tabular}[c]{@{}c@{}}Digital Relay\\ Attack\end{tabular} & \begin{tabular}[c]{@{}c@{}}Playback\\ Attack\end{tabular}           \\ \midrule
\multirow{4}{*}{Rank}                           & 1                           & \textbf{SNR}                                                       & \textbf{Kurtosis}                                              & \textbf{$f_{peak}$}                                            & \textbf{\begin{tabular}[c]{@{}c@{}}Spec.\\ Brightness\end{tabular}} \\ \cmidrule(l){2-6} 
                                                & 2                           & Kurtosis                                                           & SNR                                                            & Kurtosis                                                       & Kurtosis                                                            \\ \cmidrule(l){2-6} 
                                                & 3                           & \begin{tabular}[c]{@{}c@{}}Spec.\\ Brightness\end{tabular}         & \begin{tabular}[c]{@{}c@{}}Spec.\\ Brightness\end{tabular}     & \begin{tabular}[c]{@{}c@{}}Spec.\\ Brightness\end{tabular}     & $f_{peak}$                                                          \\ \cmidrule(l){2-6} 
                                                & 4                           & $f_{peak}$                                                         & $f_{peak}$                                                     & SNR                                                            & SNR                                                                 \\ \bottomrule
\end{tabular}
}
\end{table}

\subsection{Advanced Dual-Band Relay Attacks}\label{subsec:advc_dual}
\begin{figure}[!t]
    \centering
        \includegraphics[clip,width=0.65\columnwidth]{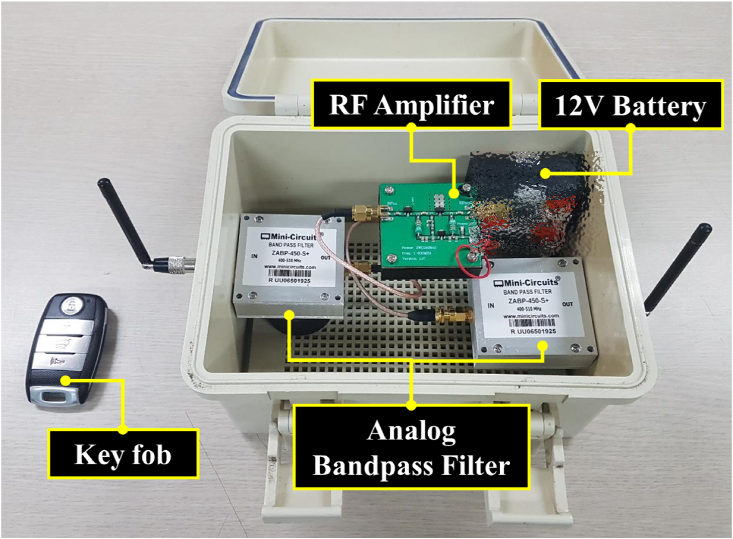}%
        \caption{Experimental setup for amplification attack simulation using pre/post-analog filter on the key fob side}
    \label{fig:amp_filter_setup}
\end{figure}

\begin{figure}[!t]
    \centering
        \includegraphics[clip,width=\columnwidth]{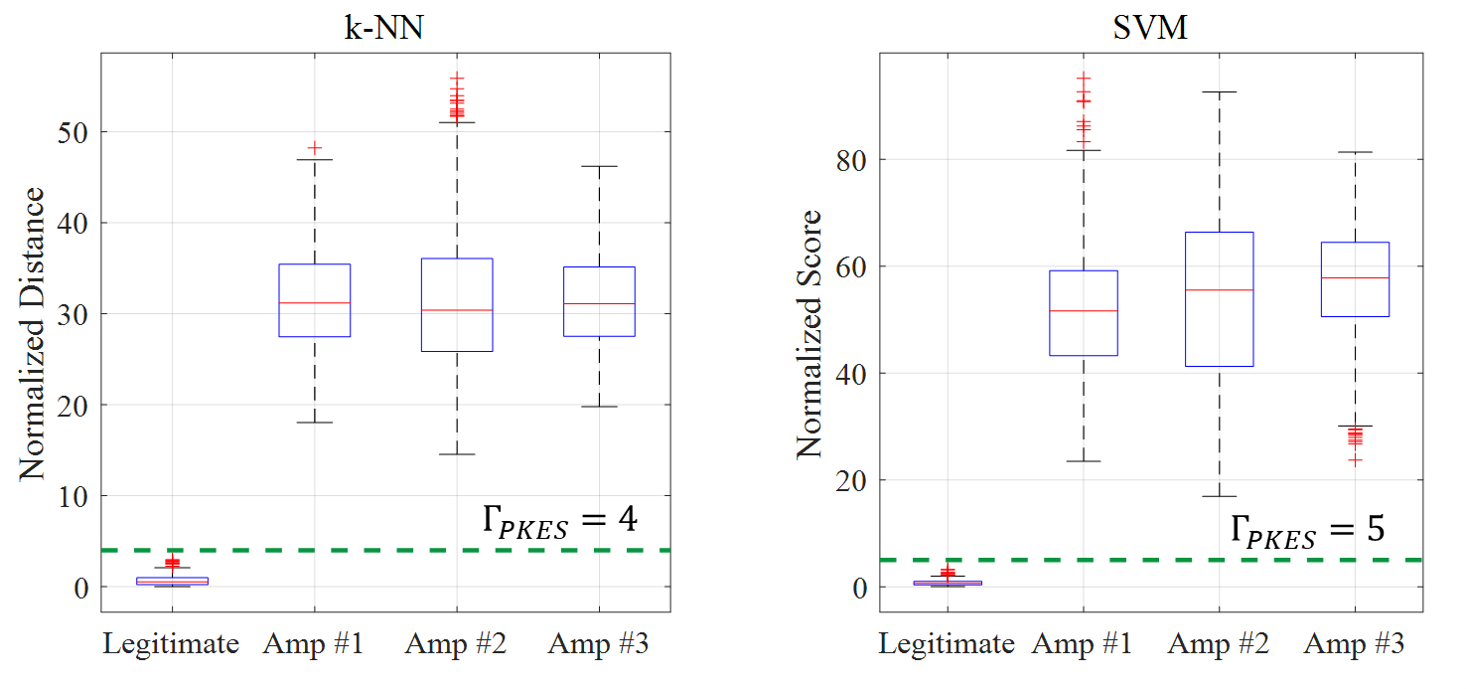}%
        \caption{Output distribution of the k-NN and SVM algorithms as a function of an amplifier under the advanced amplification attack scenario}
    \label{fig:amp_filter}
\end{figure}

\textcolor{black}{In this subsection, we assumed the presence of an advanced dual-band relay attack adversary, extending the previous experimental setup in Section \ref{subsec:dual-relay}. Advanced adversaries are equipped with analog filters or an SDR device with a high sample rate.\\
\subsubsection{Amplification attacks using analog filters}
Advanced dual-band relay adversaries can attempt to make a low value for kurtosis by using an amplifier with a pre- and post-analog bandpass filter (BPF). In this attack scenario, adversaries use pre- and post- analog filters to reduce unwanted noise. An analog filter is a circuit made of analog components such as resistors, capacitors, inductors, and op amps. Digital filters are often embedded in a chip that operates on digital signals, such as an MCU or DSP. Analog filters are fairly simple but increase in complexity for narrow bandwidth and precise roll-off. In contrast, digital filters, which are employed in \texttt{HODOR} can be more precise in filtering, but the signal must be digitally sampled. Among the bandpass filters of MiniCircuits, we used the one with the bandwidth most relevant to the UHF band \cite{ZABP450S1:online}. The low and high cut-off frequencies of the bandpass filter are 400MHz and 510MHz, respectively. Fig. \ref{fig:amp_filter_setup} shows the amplifier used with the BPF in our evaluation. The pre-/post-bandpass filters were connected to the input/output port of each amplifier. We conducted the same experiment as the one used in the amplification attack. Fig. \ref{fig:amp_filter} shows the output distributions of the k-NN and SVM algorithms as a function of the amplifiers. As seen in the figure, compared to the amplification attack without analog filters (Fig. \ref{fig:dual_relay_amp}), the normalized output distance/score is slightly closer to the legitimate case. Nevertheless, the k-NN and SVM algorithms both output an FPR of 0\% and FNR of 0\% at thresholds ($\Gamma _{PKES}$) of 4 and 5, respectively. Since the bandpass filter has a larger bandwidth (110MHz) than \texttt{HODOR} (30kHz), filtered noise has a negligible effect on the performance of \texttt{HODOR}. This is because the analog bandpass filter has a larger bandwidth than the digital bandpass filter of \texttt{HODOR}, which still amplifies the noise signals within the bandwidth of the digital bandpass filter and corrupts the value of kurtosis. In addition, if adversaries use the digital bandpass filter with a narrower bandwidth than that of \texttt{HODOR}, it accompanies both the ADC and DAC processes which corrupt the kurtosis and spectral brightness. This corruption can be observed in the evaluation results of the \textit{playback attack} detection. As a result, we concluded that \texttt{HODOR} is able to detect an amplification attack even with pre-/post-analog filters.}

\subsubsection{Playback attack with high sample rate}
\begin{figure}[!t]
  \centering
  \includegraphics[clip,width=\columnwidth]{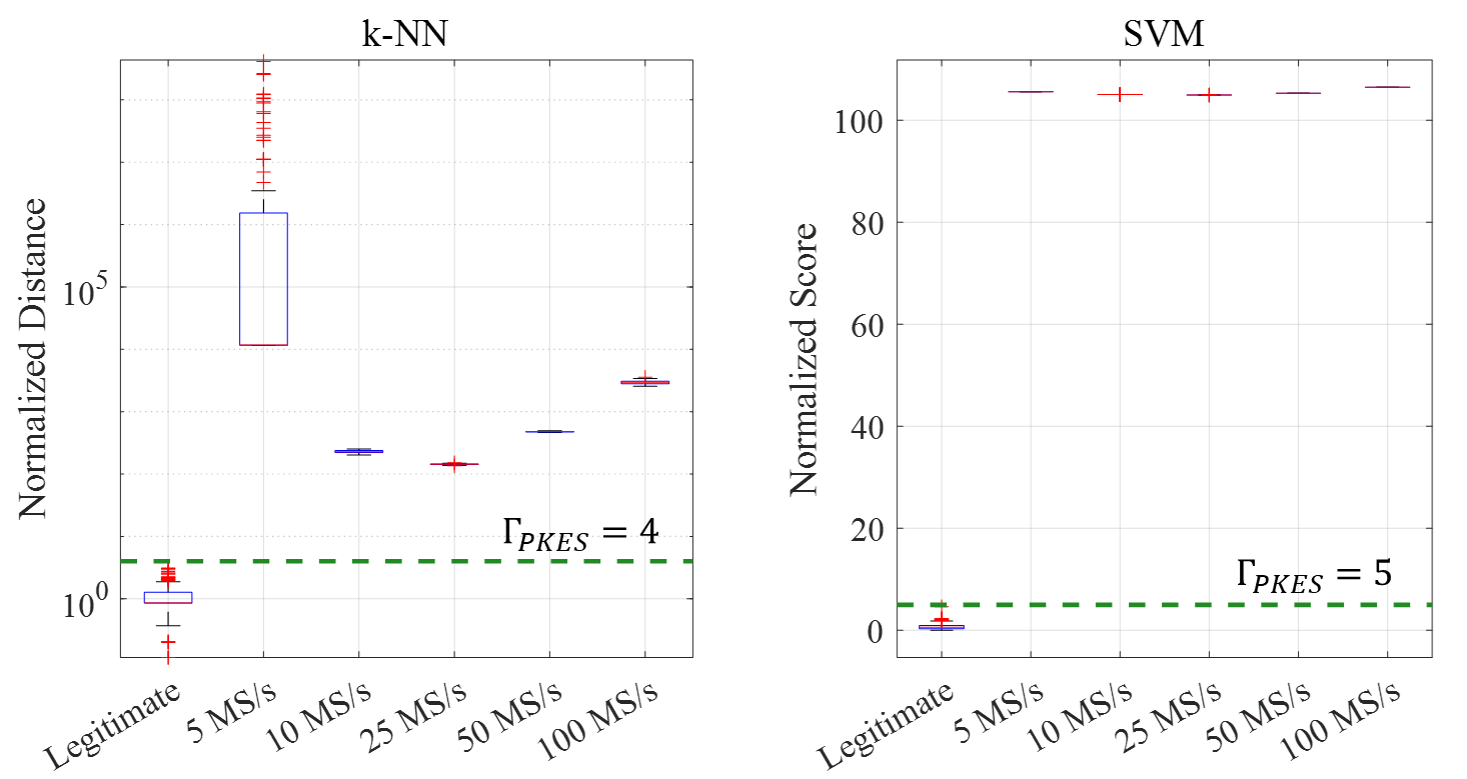}%
  \caption{Output distribution of k-NN and SVM algorithms as a function of sample rate}
  \label{fig:playback_PCIe}
\end{figure}

\textcolor{black}{
We also assumed the presence of an adversary with strong PHY-layer sampling capabilities in addition to the amplification adversaries using the analog filter. To evaluate \texttt{HODOR} with a high sample capability, we used the maximum sampling rate of the USRP X310 used in our evaluation. Because the Ethernet interface did not support the maximum sample rate, we employed a new PCIe x4 interface that can tolerate 200M samples/s on baseband signals \cite{PCIExpre81:online}. Through the PCIe X4 interface, the USRP X310 was connected with a desktop where an Intel i7-7700 3.6GHz processor CPU with 16G RAM and a 512GB SSD were installed.\\
\indent We generated UHF-band signals with high samples by using the USRP X310. \texttt{HODOR} captured the generated signals with a sample rate of 5MS/s by using HackRF One. After that, \texttt{HODOR} analyzed the features of the captured signals to compare the differences with those from a legitimate key fob. Considering the 200MHz of master clock rates (MCR) supported on the USRP X310, its sample rates must be an integer decimation rate of the MCR \cite{X300X31026:online}. Accordingly, we performed experiments with sample rates of 10MS/s, 25MS/s, 50MS/s, and 100MS/s. Theoretically, 200MS/s is also possible on the USRP X310, but the signals with a sample rate of 200MS/s cannot be properly generated by our experimental setup. It seems that higher performing hardware is required. Fig. \ref{fig:playback_PCIe} shows the output distribution of the \textit{attack with a high sample rate}. As shown in the Fig. \ref{fig:playback_PCIe}, the attacks with a higher sample rate up to 25MS/s have a much closer output distance to those with legitimate signals. Rather, the output distance increases as the sample rate exceeds 50MS/s. We conclude that a higher sample rate is not helpful to impersonate a legitimate signal. However, it should be noted that this result is from an experiment done with a software-defined radio (SDR) device. For further research on high sample rate attacks, more state-of-the-art equipment is required, such as a pair of arbitrary function generators and a tunable mixer.}\label{subsub:pcie_playback}

\subsection{Dynamic Channel Conditions}
\begin{figure}[!t]
\centering
\subfloat[]{%
  \includegraphics[clip,width=0.495\columnwidth]{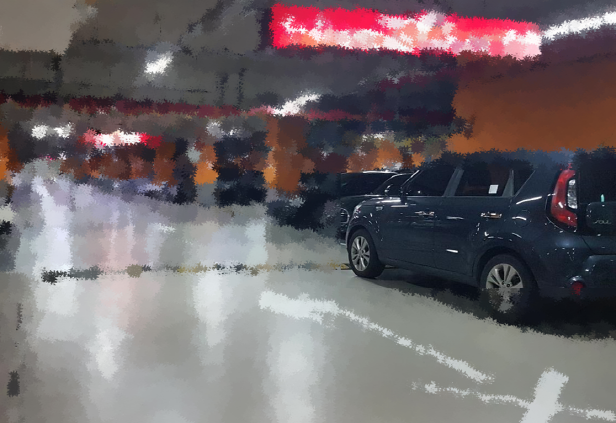}%
  \label{fig:ex_setup_underground}
}
\subfloat[]{%
  \includegraphics[clip,width=0.495\columnwidth]{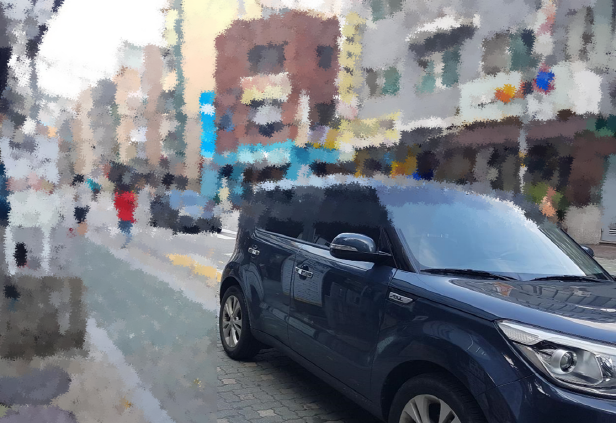}%
  \label{fig:ex_setup_roadside}
}
\caption{Dynamic environment: (a) The vehicle parked in an underground parking lot, (b) The vehicle parked in a street parking space}
\label{fig:ex_setup_busy}
\end{figure}

\begin{figure}[!t]
  \centering
  \includegraphics[clip,width=\columnwidth]{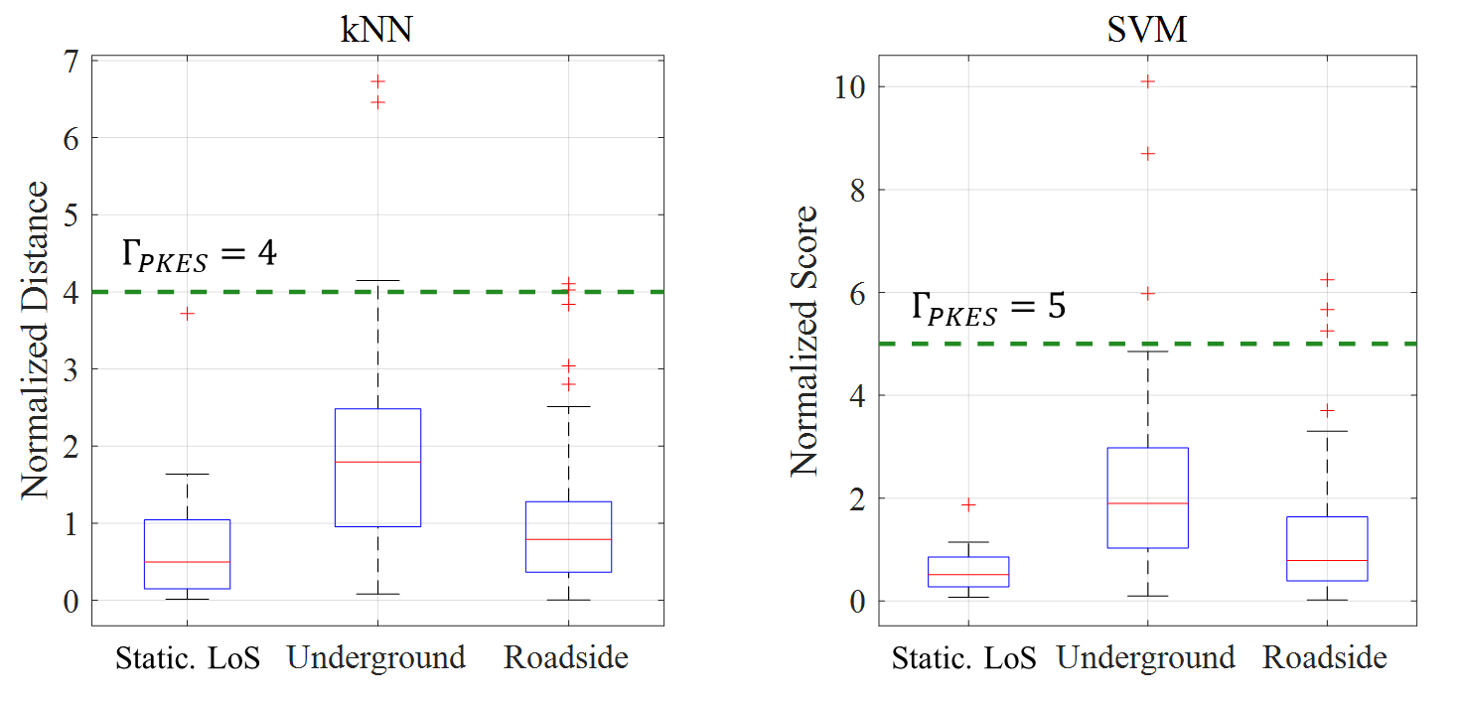}%
  \caption{Output distribution of k-NN and SVM algorithms as a function of parking space}
  \label{fig:result_nlos_busy}
\end{figure}

\textcolor{black}{A dynamic condition may decrease the performance of \texttt{HODOR} in identifying a legitimate key fob. We further evaluated \texttt{HODOR} under additional conditions, such as an underground parking lot and a roadside parking space as shown in Fig. \ref{fig:ex_setup_busy}. The underground parking lot had enough spaces for approximately 700 cars, and many cars were frequently entering and leaving the facility. Compared with an outdoor parking lot, it is expected that a greater number of multipath components are present because of the ceiling and thick concrete pillars, which can affect the performance of \texttt{HODOR}. In addition, we evaluated \texttt{HODOR} when the vehicle was parked on the road. This place was very crowded with pedestrians and moving vehicles. Furthermore, our evaluation was conducted during the most crowded hours of the day. For both environments, we placed the key fob in a backpack.
Fig. \ref{fig:result_nlos_busy} shows the output distribution of the k-NN and SVM algorithms as a function of place where the vehicle was parked. When the vehicle is parked in the underground parking lot, the k-NN and SVM algorithms output an FPR of 5$\%$ and 4$\%$ with an FNR of 0$\%$. When the vehicle is parked in a roadside parking space, the k-NN and SVM algorithms output an FPR of 2$\%$ and 3$\%$ with an FNR of 0$\%$. We should note that we used the same values for the threshold, $\Gamma _{PKES}$, as used in other evaluations. 
As can be seen in Fig. \ref{fig:nlos_static}, \texttt{HODOR} displays weaker performance in dynamic environments than in static environments. It seems that interference due to multipath components affects the features that are analyzed in \texttt{HODOR} even though their RF signals originate from a legitimated key fob. However, we believe that this result is significantly meaningful, as it demonstrates that \texttt{HODOR} is able to identify a legitimate key fob even within a dynamic environment. Finally, we further evaluated feature importance under NLoS channel conditions, including dynamic environments. In the results, the \textit{peak frequency} is the most salient feature followed by \textit{kurtosis}, \textit{SNR}, and \textit{spectral brightness}, in that order.}

%-------------------------------------------------------------------------------
\section{Related Works}
%-------------------------------------------------------------------------------

In this section, we introduce existing demonstrations of attacks on keyless entry systems and device fingerprint techniques that have been proposed in previous studies.

\subsection{Attacks on Keyless Entry Systems}
Early keyless entry systems were designed without security considerations. More specifically, RKE systems deployed in Mercedes-Benz vehicles manufactured in the early 2000s still used fixed codes \cite{garcia2016lock}. This means that these vehicles were vulnerable to replay attacks. To prevent replay attacks, a rolling code system using cryptography and a counter was installed in the RKE system. Although the rolling code provides an adequate level of security, several studies \cite{indesteege2008practical, eisenbarth2008power} demonstrated the cryptographic weakness of the KEELOQ system via a cryptographic analysis. For example, Garcia et al. \cite{garcia2016lock} performed a comprehensive survey and security analysis of the RKE system implemented in vehicles manufactured by Volkswagen. The authors describe various types of RKE systems used in Volkswagen vehicles, and revealed that they were using a single worldwide secret key. Moreover, Benadjila et al. \cite{benadjila2017one} demonstrate attacks on RKE systems using the Hitag-2 cryptographic algorithm. The authors demonstrate that it is possible to extract the rogue key (a.k.a. not the true secret key), while, in turn, exploiting a weakness of the Hitag-2 cryptographic algorithm. More recent studies further reveal a weakness of the AUT64 automotive cryptographic algorithm. In \cite{hicks2018dismantling}, the authors demonstrate the full details of AUT64 including a complete specification and analysis of a block cipher. The authors propose two key-recovery attacks based on cryptographic weakness. In addition to studies based on the weaknesses of cryptographic algorithms, roll jam attacks on RKE systems have also been studied \cite{yang2018inside}. In this type of attack, an adversary avoids being discovered by the victim while jamming and eavesdropping, and at the same time, obtains a valid rolling code. In this scenario, the car owner would not suspect an attack despite having pressed the button on their key fob twice to unlock the doors.\\
\indent Alrabady et al. \cite{alrabady2003some} also analyze potential attacks on PKES systems. The authors demonstrate several potential attacks on passive entry systems and propose solutions to protect vehicles from such attacks. However, to deploy these proposed solutions, the frequency band of the key fob and the vehicle would have to be altered, which forces the previous PKES system to be re-designed. The same authors also analyzed various types of security attacks against keyless entry systems and compared the attacks in terms of the vulnerability of the security mechanism deployed on the system, level of difficulty to mount the attacks, and equipment needed for the attacks \cite{alrabady2005analysis}. Francillon et al. \cite{francillon2011relay} pragmatically show a different type of attack on PKES systems, the so-called relay attack. The authors performed a comprehensive experiment and proposed two types of attack scenarios: wired and wireless relay attacks. In addition, \cite{unicornteam:online} showed a similar but still different type of relay attack on PKES systems. In this work, adversaries were presumed to be extracting binary codes from RF signals and relaying these to another location. Although there is a maximum receiving time delay allowance in the PKES system, the authors showed that it was possible to overcome this constraint using cheap devices. Finally, in real car theft cases \cite{Keylesss80:online, bbctehft:online}, it was revealed that the adversaries amplified both the LF-band and UHF-band signals to mount a relay attack known as an amplification attack.
\vspace{-0.2cm}

\subsection{RF Device Fingerprinting}
Depending on the signal region used to extract fingerprints, approaches can be categorized as transient-based, modulation-based, or other approaches \cite{danev2012physical}. Transient-based approaches use the turn-on/off transient of an RF signal for device identification. This approach has been applied to the identification of VHF FM transceivers or intrusion detection in a WLAN environment \cite{ellis2001characteristics, hall2005radio}. However, transient signals have been reportedly acquired at distances close  to the front-end antenna (10 $\sim$ 20cm),  and this is unfeasible in keyless entry systems. Modulation-based approaches extract features from the base band signal. In this study, the classifier achieves a classification error rate of 3$\%$ and 0.34$\%$ for the k-NN and SVM classifiers, respectively. However, to achieve this error rate, a high-end vector signal analyzer \cite{VectorSignalAnalyzers:online} is necessary for signal acquisition.\\
\indent Other approaches considered other regions of the RF signals to extract the fingerprints, such as the preamble of the packet. As such, Suski et al. \cite{suski2008using} proposed a power spectrum density of the packet preamble to extract fingerprints. Jun et al. \cite{huang2014blueid} also presented BlueID, which fingerprints the clock of a Bluetooth device. BlueID estimates the clock skew of a Bluetooth device applying the Generalized Hough Transform (GHT). In this work, 200 received preambles are required even though the time-stamp value is not required to calculate a clock skew. In a keyless entry system, however, the maximum five preambles can be received for attack detection. \textcolor{black}{Recently, the authors of \cite{chatterjee2018rf} proposed an RF-PUF which exploits the unique hardware characteristics of an RF device to an authenticated device. The main concept of \texttt{HODOR} is similar to that of an RF-PUF, given that both analyze RF signals to develop a security method, such as authentication or identification. However, in contrast to \texttt{HODOR}, multiple wireless devices used in an RF-PUF were simulated based on software implementation. \texttt{HODOR} was evaluated using physical hardware rather than simulations in environments while considering possible scenarios in the PKES system.}
%\vspace*{-0.3cm}
%%% 여기까지
%-------------------------------------------------------------------------------
\section{Discussion}
%-------------------------------------------------------------------------------
\vspace{-0.1cm}
\subsection{\texttt{HODOR} and Security}
\textcolor{black}{A threshold is a trade-off parameter in \texttt{HODOR}. If the output score of a UHF-band signal and its features is larger than the threshold, \texttt{HODOR} would determine that the signal is not from a legitimate key fob. In addition, \texttt{HODOR} is designed with a sufficiently large threshold so that it can tolerate some degree of error. For example, the environments under an NLoS channel and high-temperature variation cause a high noise level, and \texttt{HODOR} should tolerate a feature variation that occurs under these conditions. Because of this large threshold, on the other hand, \texttt{HODOR} might accept a UHF-band signal even if its features are not perfectly impersonated. We analyzed the difficulties of feature impersonation to additionally discuss \texttt{HODOR}'s security in the following subsection.}
\vspace{-0.15cm}
\subsection{Feature Impersonation}
\textcolor{black}{Difficulty of feature impersonation is another crucial factor in evaluating the security level of \texttt{HODOR}. Since \texttt{HODOR} employs multiple features for attack detection, the degree of feature impersonation difficulty should be discussed separately.}\par
\textbf{Peak frequency. }\textcolor{black}{The peak frequency represents the bit time characteristics of an individual device. In a digital relay attack and cryptographic attack, the peak frequency was clearly different between attack signals and legitimate signals. Because every RF device has a different bit time, it is possible to identify RF devices based on their bit time. Moreover, the peak frequency enables \texttt{HODOR} to identify RF devices even if they have identical components with each other. On the other hand, the playback attack that records and playbacks UHF-band signals from a legitimate key fob is able to similarly impersonate the peak frequency. However, even though the peak frequency was successfully impersonated by the playback attack, it fails to impersonate the spectral brightness and kurtosis which are affected by the ADC and DAC process.}\par
\textbf{SNR. }\textcolor{black}{Even though the SNR feature is effective in detecting single-band relay attacks, an adversary can impersonate the SNR of a legitimate signal by using an amplifier or by varying the distance from the vehicle. However, the both methods also affect kurtosis similar to the ways to impersonate the peak frequency feature. Using an amplifier especially increases noise level as well as the baseband signal level. Due to this amplified noise level, it is difficult to impersonate kurtosis while also impersonating SNR. In addition, channel condition between the two adversaries affect the kurtosis feature. An ideal analog filter would perfectly amplify only the baseband signal level, and not noise level. However, it is very difficult to design the ideal analog filter in practice because of the analog circuit complexity.}\par
\textbf{Spectral Brightness. }\textcolor{black}{The spectral brightness feature represents the amount of energy in a high-frequency region of signals. This feature is very helpful to detect attacks where new ADC and DAC processes occurs, such as in a playback attack. However, the spectral brightness of each key fob has a similar value. As a result, the attacks must be executed with identical components as the target key fob in order to impersonate the spectral brightness feature. However, different devices have different bit times that are represented by the peak frequency feature.}\par
\textbf{Kurtosis. }\textcolor{black}{The kurtosis feature was the most salient feature across the series of evaluations. Although the kurtosis feature is not helpful to detect the digital relay attack using identical components (i.e., other key fobs), it outperforms when the single-band relay attack and amplification attack are detected. However, the kurtosis feature of each key fob is similarly distributed as spectral brightness. This implies that an attack device with identical components as a target key fob is required to impersonate its kurtosis feature. Even if two devices are designed with identical components, as mentioned above, they have different bit times with each other, which are represented by peak frequency. Accordingly, if the kurtosis feature is successfully impersonated by using a device with identical components, its attack would be detected based on the different peak frequency of an attack signal.}
\vspace{-0.15cm}
\subsection{Concern for Practicality}
\textcolor{black}{Although we evaluated \texttt{HODOR} under various environmental conditions, some results seem still insufficient for the practical usage of \texttt{HODOR}. It has been shown that \texttt{HODOR} has a relatively high FPR under dynamic conditions and temperature variation. Other than the conditions where we evaluated \texttt{HODOR}, more extreme environments certainly do exist that adversely affect accuracy. For this reason, \texttt{HODOR} should be further studied to develop additional features and algorithms that properly operate even in extreme environments. Subsequently, other processes will be performed together with \texttt{HODOR} to unlock doors. Regarding the total execution time for a door unlock command, \texttt{HODOR} should also be improved to shorten execution time. Otherwise, a driver may sense a lag when he/she attempts to unlock the doors. As a result, \texttt{HODOR} should be further studied to resolve these practical issues. Through our future research, it is expected that \texttt{HODOR} will improve and achieve the requirements for practical usage.}

\subsection{Scalability} \textcolor{black}{The concept that analyzes RF signals can be applied to other modulations for secure proximity and distance measurement. However, such analysis is beyond the scope of this paper and further research on the characteristics of baseband signals generated by other modulations is required. We suspect that \texttt{HODOR} may be applicable in other modulations for secure proximity and distance measurement since the hardware characteristics of an attack device and channel conditions affect the features.}
\vspace{-0.15cm}
\section{Conclusion}
In this paper, we presented \texttt{HODOR} as a sub-authentication system that supports manufacturer-installed support systems to prevent keyless entry system car theft. \texttt{HODOR} is an RF-fingerprinting method that distinguishes a legitimate door unlock request from a malicious attempt. Through our evaluation, we showed that \texttt{HODOR} is able to effectively detect simulated attacks that are defined in our attack model, while reducing the number of erroneous detection occurrences (i.e., false alarms). Furthermore, we found a set of suitable features in a number of environmental conditions, such as temperature variation, battery aging, and NLoS conditions, that make it possible for \texttt{HODOR} to properly operate in real-life environments. Finally, one especially noteworthy merit of \texttt{HODOR} is its design. It is designed such that it can be applied into an existing system without any hardware modifications. The only requirement for successful implementation is to add a device to sample UHF-band RF signals and analyze them. This novel characteristic of our method means that \texttt{HODOR} improves security without creating additional cumbersome or inconvenient processes for the user.

\section*{Acknowledgments}
The authors would like to thank the anonymous reviewers for their valuable feedback. This work was supported by Samsung Research Funding $\&$ Incubation Center for Future Technology under Project Number SRFC-TB1403-51.

\ifCLASSOPTIONcaptionsoff
  \newpage
\fi

\bibliographystyle{IEEEtranS}
\bibliography{IEEEabrv,refer}

\end{document}